\documentclass[twocolumn,tighten]{aastex61}
\usepackage{color}

\newcommand{\ie}{i.e.,\ }
\newcommand{\eg}{e.g.,\ }
\newcommand{\etal}{et~al.\ }
\newcommand{\kms}{km~s$^{-1}$}
\newcommand{\magsec}{mag arcsec$^{-2}$}
\newcommand{\mv}{F606W}

\newcommand{\mi}{F814W}
\newcommand{\Mi}{M$_{\rm F814W}$}
\newcommand{\vmi}{F606W~$-$~F814W}

\begin{document}

\title{Stellar Populations in the Outer Disk and Halo of the Spiral Galaxy M101}
\shorttitle{Stellar Populations in the Outskirts of M101}

\author{J. Christopher Mihos}
\affiliation{Department of Astronomy, Case Western Reserve University,
Cleveland, OH 44106, USA}

\author{Patrick R. Durrell}
\affiliation{Department of Physics and Astronomy, Youngstown State University,
Youngstown, OH 44555, USA}

\author{John J. Feldmeier}
\affiliation{Department of Physics and Astronomy, Youngstown State University,
Youngstown, OH 44555, USA}

\author{Paul Harding}
\affiliation{Department of Astronomy, Case Western Reserve University,
Cleveland, OH 44106, USA}

\author{Aaron E. Watkins}
\affiliation{Department of Astronomy, Case Western Reserve University,
Cleveland, OH 44106, USA}
\affiliation{University of Oulu, Astronomy Research Unit, FI-90014 Oulu, Finland}

\begin{abstract}

We use deep {\sl Hubble Space Telescope} imaging in the outskirts of the 
nearby spiral M101 to study stellar populations in the galaxy's outer disk
and halo. Our ACS field lies 17.6\arcmin\ (36 kpc) from the center of
M101 and targets the blue ``NE Plume'' of M101's outer disk, while the parallel WFC3 
field lies at a distance of 23.3\arcmin\ (47 kpc) to sample the galaxy's stellar
halo. The WFC3 halo field shows a well-defined red giant branch characterized
by low metallicity ([M/H]~$=-1.7 \pm 0.2$), with no evidence of young
stellar populations. In contrast, the ACS disk field shows multiple stellar
populations, including a young main sequence, blue and red helium burning
stars, and old RGB and AGB populations. The mean metallicity of these disk stars
is quite low: [M/H]~$=-1.3 \pm 0.2$ for the RGB population, and $-1.15 \pm 0.2$
for the younger helium burning sequences. Of particular interest is a bunching
of stars along the BHeB sequence, indicative of an evolving cohort of massive
young stars.  We show that the young stellar populations in this field are well-described 
by a decaying burst of star formation that peaked $\sim$ 300--400 Myr ago,
along with a more extended star formation history to produce the older RGB and
AGB populations. These results confirm and extend the results from our previous 
deep surface photometry of M101's outer disk, providing an important cross-check
on stellar population studies using resolved stellar populations versus integrated
light photometry. We discuss our results in the context of halo formation models
and the interaction history of M101 and its companions.

\end{abstract}


\section{Introduction}

The diffuse outskirts of disk galaxies are shaped by a wide variety of
physical processes linked to the evolution of galaxies. Ongoing,
low-level star formation --- as suggested by the extended ultraviolet
(XUV) disks seen in many spiral galaxies (\eg Gil de Paz \etal 2005,
Thilker \etal 2007, Lemonias \etal 2011) --- can gradually build the
stellar populations over time. Interactions and accretion can deposit
stars into the outer disk and halo (\eg Bullock \& Johnston 2005,
Villalobos \& Helmi 2008, Martinez-Delgado \etal 2008, 2009 Cooper \etal
2010, Ruiz-Lara \etal 2016), or trigger new star formation in the disk
outskirts (\eg Werk \etal 2010, Mihos \etal 2013, Bush \etal 2014).
Tidal heating by galaxy interactions can also drive metal-rich stars
outwards from the inner disk (Walker \etal 1996, Younger \etal 2007),
while secular processes such as radial migration can scatter inner disk
stars outwards as well (Sellwood \& Binney 2002, Debattista \etal 2006,
Roskar \etal 2008ab). These processes imprint signatures on the spatial
distribution, kinematics, ages, and metallicities of the outer disk
stellar populations that can be used to reconstruct the evolutionary
history of the outer disk.

However, studying the stellar component of galaxy outskirts has proved
particularly challenging. While the distribution and kinematics of the
extended HI disk can be probed relatively simply via 21-cm line emission
mapping, the surface brightness of the outer {\sl stellar} component
($\mu_V \gtrsim 26$ \magsec) is well below that of the night sky,
requiring very deep and accurate surface photometry to measure its
structure and optical colors (\eg Bakos \etal 2008, Erwin \etal 2008, Watkins \etal 2014,
2016, Zheng \etal 2015, Merrit \etal 2016, Peters \etal 2017). A
complementary approach is available for nearby galaxies ($\lesssim$10
Mpc): the study of the discrete stellar populations via deep {\sl Hubble
Space Telescope} (HST) imaging of individual stars in a galaxy's disk.
The distribution of stars along the color-magnitude diagram (CMD) is
sensitive to the age and metallicity of the underlying stellar
population, and can be used to trace the star forming history and
metallicity evolution of the galaxy (\eg Rejkuba \etal 2005, Dalcanton
\etal 2009, Williams\etal 2009, 2001, Radburn-Smith \etal 2011, McQuinn
\etal 2010, 2011, 2012, Bernard \etal 2015, Bruzzese \etal 2015,
Monachesi \etal 2016).

As powerful as the discrete stellar population technique is, its ability
to probe distant galaxies is limited by by the need to actually resolve
individual stars. Ground-based studies have been restricted to
relatively nearby galaxies within the Local Volume (\eg Brooks \etal
2004, Davidge 2006, 2010, Harris \& Zaritsky 2009, Ibata \etal 2014),
HST studies have pushed as far out as the Virgo Cluster (\eg Ferguson
\etal 1998, Harris \etal 1998, Caldwell 2006, Durrell \etal 2007,
Williams \etal 2007, Bird \etal 2010), but typically only detect the
brightest RGB stars; studying the fainter populations becomes
prohibitively expensive in terms of telescope time. At even larger
distances, studies of stellar populations in galaxy outskirts are
limited to deep surface photometry, which provide much weaker
information, due to the poor discrimination broadband colors have to
ages greater than a few Gyr, coupled with the well-known age-metallicity
degeneracy. While star formation rates can be readily probed through
deep H$\alpha$ or FUV imaging (see, \eg the review by Kennicutt \& Evans
2012), these are sensitive only to timescales of $\lesssim$ 100 Myr, the
lifetime of the massive UV-producing stars. Obtaining the multicolor
imaging or spectrophotometry that yields more nuanced information over a
wider range of stellar ages is largely prohibitive at such low surface
brightness. The limited reach of discrete stellar population work
coupled with the degeneracies of integrated light techniques leaves us
with a dissatisfying seam in studies of the star forming histories of
galaxies. Much of our detailed information in the Local Group come from
discrete stellar populations, while our more general understanding of
star forming histories of galaxies comes from studying large samples of
more distant galaxies using imaging and spectrophotometry of their
integrated light. The differing systematics of these two techniques
means that connecting lessons learned from the Local Group with those
from more distant galaxies remains uncertain.

The nearby spiral galaxy M101 (NGC 5457, $D=6.9$ Mpc; Matheson \etal
2012 and references therein) provides a particularly important test case
for studying the evolution of disk outskirts and connecting deep surface
photometry studies with discrete stellar population work. M101 is the largest member
of a small galaxy group (Geller \& Huchra 1983, Tully 1988), an
environment where interactions with companions can drive significant
evolution in the galaxy's disk. Deep far UV imaging from GALEX has shown
M101 to be an XUV disk (Thilker \etal 2007), with star formation
occurring at large radius well outside the bright inner disk. M101 has
been well-studied at many wavelengths, giving us a comprehensive view of
the galaxy's structure, kinematics, and star forming properties. Its
proximity also makes it an optimal target for studying the stellar
populations of its outer disk and halo via discrete stellar population
techniques.

The complicated outer structure of M101's disk was first revealed in
deep, wide-field surface photometry of the galaxy by Mihos \etal (2013;
hereafter M+13). Beyond 12.5\arcmin\ (25 kpc), the outer disk shows two
distinct two stellar plumes --- dubbed the Northeast (NE) Plume and
Eastern Spur by M+13 --- suggestive of tidal features from a past
encounter. The NE Plume in particular showed extremely blue colors 
($B-V \approx 0.2$), but little ongoing star formation as evidenced by the
lack of far UV light. This led M+13 to propose that the outer disk had
experienced a weak burst of star formation that peaked 250--350 Myr ago
and had since largely died out. But due to the age-metallicity
degeneracy of broadband imaging, little could be said of metallicity of
the population, the robustness of the inferred star forming history, or
the presence of any underlying older populations. Aside from the
stellar plumes, little additional light was detected beyond the galaxy's
outer disk, a fact which led van Dokkum \etal (2014) to propose that
M101 had a very anemic stellar halo, down by nearly an order of
magnitude in stellar mass fraction compared to the Milky Way and M31.

Discrete imaging of M101's outskirts using HST has the potential to
address many of the issues surrounding the evolutionary history of
M101's disk and the presence of halo stellar populations. The colors and
number counts of stars of differing evolutionary phases along the CMD
--- along the main sequence, in helium burning phases, or on the red
giant or asymptotic giant branches --- will give a much more robust
measure of the star formation history of the outer disk and provide, for
the first time, metallicity estimates of the galaxy's stellar
populations. Furthermore, because star count techniques have the ability
to go to lower equivalent surface brightness than deep surface
photometry (\eg Ibata \etal 2014, Okamoto \etal 2015, Crnojevi\'{c}
\etal 2016), we can also make a direct measure of the density of M101's
stellar halo. Finally, aside from learning about the evolutionary
history of M101 itself, follow-up HST imaging of discrete populations in
its outer disk is permit valuable cross-comparisons of the deep surface
photometry and discrete stellar population imaging techniques.

In this paper, we follow up on the imaging of M+13 by studying the NE
Plume using HST ACS imaging that extends deep enough to detect not just
the RGB stars and massive helium burning stars, but the young main
sequence population as well. By constructing a deep CMD of the stellar
populations, we can use isochrone matching and CMD modeling techniques
to infer the metallicity distribution of stars in M101's outer disk, and
probe the star forming history of the stellar plumes with much finer
discriminatory power than was possible using the broad band imaging
techniques of M+13. We also use WFC3 imaging of a blank field near the
NE Plume to search for and characterize any stellar populations
associated with M101's stellar halo, to assess the conjecture that M101
has an anomalously weak halo component. Finally, we also make a direct
comparison of these results to deep ground-based surface photometry of
M101 to connect these two complementary methods for studying stellar
populations in galaxies.

\section{Observations and Data Reduction}

\subsection{Field Selection and Observations}

\begin{figure*}[]
\centerline{\includegraphics[width=7.0truein]{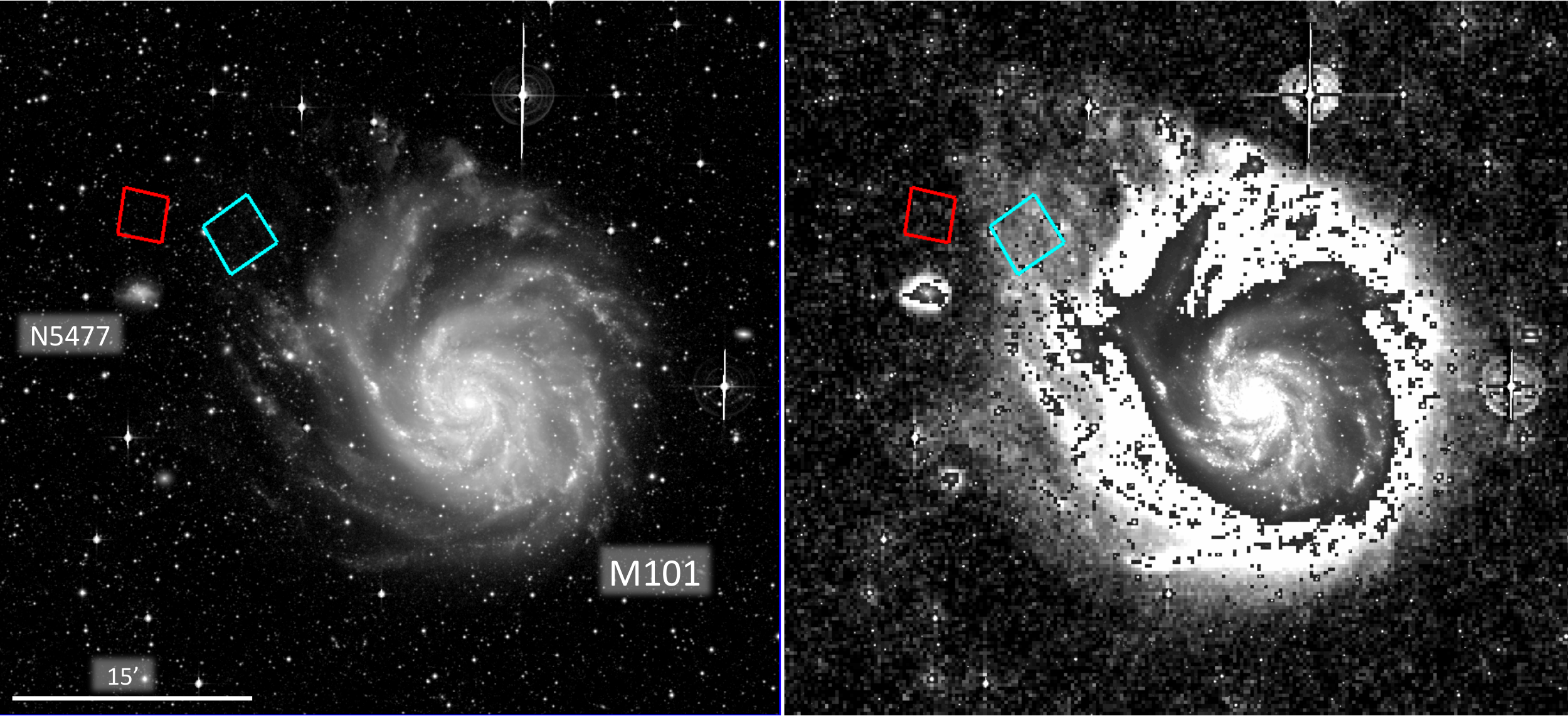}}
\caption{Our HST fields (ACS: blue; WFC3: red) overlaid on deep $B$
imaging of M101 from Mihos \etal (2013). The left panel shows the image
at 1.5\arcsec\ pixel$^{-1}$ resolution while the right image shows the
image reprocessed to show the low surface brightness outer disk; the
limiting surface brightness in the image is $\mu_B=29.5$ \magsec\ (see Mihos \etal
2013 for details).}
\label{fields}
\end{figure*}

To sample the stellar populations in M101's outer disk and halo, we
obtained observations of a pair of fields (see Figure~\ref{fields})
using the both the Wide Field Channel (WFC) of the Advanced Camera for
Surveys (ACS) and the UVIS channel of the Wide Field Camera 3 (WFC3)
on-board the $HST$, as program GO-13701. The primary ACS field (covering
$\approx3.4\arcmin \times 3.4\arcmin$) lies on the NE Plume, centered at
$\alpha=14^h 04^m 50\fs 62$, $\delta= +54\degr 31\arcmin 19.1\arcsec$
(J2000.0), with a projected distance of 17.6\arcmin\ (36 kpc) from the
center of the galaxy. The field was chosen to avoid the most actively
star-forming regions of the NE Plume, and accordingly is somewhat redder
($B-V=0.31$) than the Plume as a whole ($B-V=0.21$; M+13). The parallel
WFC3 field, located at $\alpha=14^h 05^m 31\fs 93$, $\delta= +54\degr
32\arcmin\ 26.2\arcsec$ (J2000.0), covers a region outside the disk
where no diffuse starlight was detected by M+13 down to a surface
brightness limit of $\mu_B=29.5$ \magsec. The field lies at a projected distance of
23.3\arcmin\ (47 kpc) from the center of M101 and will serve as both a
control/background field for the ACS imaging, as well as to provide a
glimpse into the stellar populations of M101's halo.

The two fields were observed over 26 orbits on 2015 Sep
21--26, with two images per orbit. The depth of our imaging
is driven by the desire to detect not only the luminous old RGB
and young helium burning stars, but also the young main sequence
population down to an age of $\sim 400$ Myr. At M101's 
distance, this age corresponds to a main sequence turnoff at 
F814W $\approx 28.5$. The primary ACS observations were
split into two orbit visits with an exposure series of (1320s, 1320s,
1415s, 1415s) per visit. The F606W observations included 7 visits, for a
total exposure time of 38290s; the F814W observations were taken in 6
visits, for a total exposure time of 32820s. The exposure time series
for the parallel WFC3 observations were (1400s, 1400s, 1450s, 1450s),
with total exposure times of 34200s for the F606W observations over 6
visits, and 39900s for the F814W observations over 7 visits --- the
longer exposures times for F814W were needed to account for the lower
quantum efficiency of the WFC3 camera in F814W. In each visit, we used a
customized 4 point sub-pixel DITHER-BOX pattern in order to maximize the
sampling of the PSF over both cameras. Each visit was dithered with 3
larger $\Delta x=20$ pixel ($\sim 1\arcsec$) offsets to effectively
remove bad pixels, with this pattern repeated at an additional $\Delta y
= 60$ pixel ($\sim 3\arcsec$) shift to cover the ACS chip gap, allowing
for a contiguous study of the stellar populations in our fields. We
retrieved the calibrated, CTE-corrected {\it .flc} images for both
cameras in May 2016, in order to make use of images with the best
possible calibration files applied.

\subsection{Point Source Photometry}

We used the ACS module of the DOLPHOT 2.0 software package (Dolphin 2000)
for the photometric analysis of the ACS field; this package is
specifically designed for point-source photometry of objects in the
individual CTE-corrected {\it .flc} images. Object detection and
photometry was performed on all exposures simultaneously, using a deep
image stack derived from multidrizzling (using drizzlepac 2.0) all 28
individual F814W images for use as the reference image. The DOLPHOT
parameters were set to the recommended values for ACS/WFC images as
given in the DOLPHOT User's Guide\footnote{http://purcell.as.arizona.edu/dolphot/}, which includes
PSF-fitting using pre-derived PSFs, as well as the calculation and
application of aperture corrections. Sky values were determined using
large regions around each object, as the images are not crowded and have
no large-scale background sky variations. The instrumental magnitudes
were converted to the VEGAMAG $HST$ photometric system by adopting 
updated F606W and F814W zeropoints for 2016 data products.

Photometric reductions on the individual WFC3 images were conducted with
DOLPHOT in a similar way as the ACS reductions. Jang \& Lee (2015),
however, note small differences ($\lesssim 0.05$ mag) between the F606W
and F814W VEGAMAG values between the two cameras. Thus, to facilitate a
cleaner comparison between the two datasets, we have applied their
conversions to put the WFC3 magnitudes on the ACS system. We refer to
magnitudes in the ACS VEGAMAG system by their filter name (\mv\ and
\mi), and in cases when we need to convert to the Johnson-Cousins
systems we refer to those filters as $V$ and $I$.

We determined the photometric completeness limits and associated errors
in our point-source photometry by adding and remeasuring 100,000
artificial stars in each of the ACS and WFC3 image sets with DOLPHOT.
The input artificial stars had a wide range of magnitude
(24.0~$<$~F606W~$<$~31.0) and color ($-1.0<$~F606W$-$F814W~$<2.5$) in
order to investigate completeness for all of the possible stellar
populations in our CMDs. In processing the artificial star photometry,
we also apply the same photometric cuts on detection threshold and image shape that we
use on the real data (see next section), to ensure the best measure
of photometric errors and completeness in our extracted CMDs.
In the ACS imaging, we find a 90\% (50\%)
completeness limit of 28.6 (29.4) in \mv\ and 27.8 (28.6) in \mi. For
the WFC3 imaging, the 90\% (50\%) limits are 28.6 (29.5) in \mv\ and
27.7 (28.6) in \mi. We also use the artificial star tests to determine
photometric uncertainty, with characteristic errorbars as a function of
magnitude shown in Figure~\ref{photcrit}. At faint magnitudes, near the
50\% completeness limits, we find a slight blueward shift of $\sim 0.06$
mag in the \vmi\ colors, illustrated by the shift of the
Figure~\ref{photcrit} errorbars off the \vmi~$=-1$ axis line. However,
this shift drops to less than 0.01 mag at brighter magnitudes
(\mi~$<26.5$), and at all magnitudes is always much smaller than the
photometric uncertainty.

\subsection{Color-Magnitude Diagrams}

To create the F606W, F814W color-magnitude diagrams (CMDs) for each of
our fields, we first applied a series of criteria in order to extract
the best possible photometry for those point sources detected in both
the F606W and F814W images. Only those objects with DOLPHOT object
TYPE$=1$ (``good star'') and combined S/N values $>3.0$ in {\it both}
filters were retained for further analysis. While these cuts remove the
lowest S/N photometry and obvious non-stellar objects, we used
additional criteria from the DOLPHOT output to remove as many
non-stellar objects as possible, while keeping those point sources with
the best possible photometry. To this end, we retained only those
objects that were measured on at least 8 of the images in both filters
($N_{F814W}>8$, $N_{F606W}>8$), those with $CROWD<0.4$ mag, and
goodness-of fit $CHI<2.0$ in both filters. Finally, we used only those
objects with sharpness values consistent with point sources,
$|sharp|<0.08+0.25 e^{(F814W-28.5)}$. This function was used to select
only point sources, and is similar to that used to successfully identify
M81 halo star populations in Durrell \etal (2010). The best choice
coefficients for the function were determined empirically through visual
inspection of objects on the drizzled/stacked F606W and F814W images,
down to $F814W=27$. To ensure that we are not being overly
aggressive and removing true point sources, we examine the effect of
this criterion on our artificial star tests, where we find that fewer
than 1\% of the artificial stars brighter than our 50\% completeness
limits are being rejected by this choice.

\begin{figure}[]
\centerline{\includegraphics[width=3.5truein]{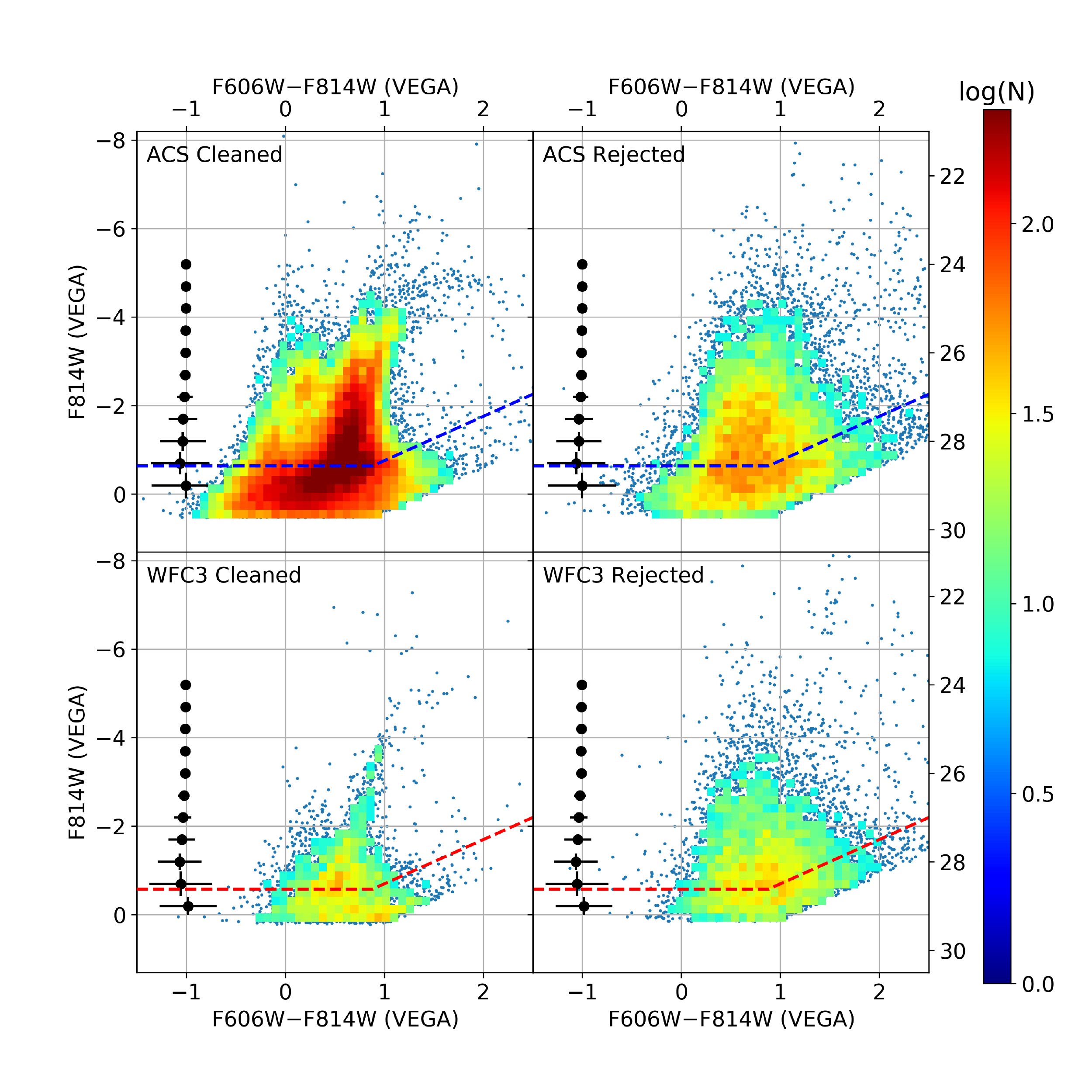}}
\caption{The photometrically cleaned stellar sources (left panels) and
the rejected objects (right panels) plotted on the F606W, F814W color
magnitude diagram. The CMDs show no structure in the distribution of
rejected sources that would indicate over-rejection of stellar sources
in M101. Characteristic errorbars are shown in each CMD, while the
dashed lines show the 50\% completeness depth of the data. In these and
all subsequent CMD plots, the right axis shows the observed apparent
magnitude of the objects, while the left axis shows the absolute
magnitude of the source at the M101 distance (6.9 Mpc).}
\label{photcrit}
\end{figure}

The final, cleaned CMDs for each of the ACS and WFC3 fields (after
application of all criteria above) are presented in
Figure~\ref{photcrit}. One potential concern is that our rejection of
non-stellar sources may be {\sl too} aggressive, and eliminate bona-fide
M101 stars in addition to the contaminating sources. To test this,
Figure~\ref{photcrit} also shows the CMDs for objects that did {\sl not}
pass the sharpness, $CROWD$, $CHI$, $N_{F606}$ and $N_{F814}$ criteria.
If we are overly aggressive in our rejection of non-stellar sources, we
should see stellar populations sequences (such as RGB or helium burning
stars) in the rejected-source CMDs. However, inspection of these
rejected objects shows no obvious population sequences and very little
structural similarity to the cleaned stellar-source CMDs, giving us
confidence that we are not overly rejecting M101 stars in our
photometric cleaning.

After all cuts, the ACS CMD has a total of 33,967 sources in it,
with 8,924 (17,417) sources brighter than the ACS 90\% (50\%)
completeness limit. The WFC3 CMD has a total of 3,513 sources, with
590 (1,786) sources brighter than the WFC3 90\% (50\%) completeness
limit.

\section{Background Source Estimation}

Given the depth of our observations, there is potential for significant contamination
both from foreground Milky Way stars and unresolved background galaxies. Following the
work of others (\eg Radburn-Smith et al. 2011; Dalcanton et al.
2012 ; Monachesi et al. 2016), we first estimated foreground Milky Way
contamination. Using both the
Besan\c{c}on (Robin \etal 2003)
and TRILEGAL (Girardi \etal 2005, 2016) models
in their default configurations, we simulated one square degree of Milky
Way sources around the central coordinate of our target locations to
ensure well-populated CMDs. We then rescaled the counts to the actual
area of the ACS and WFC3 fields, and used the photometric transformation
of Sirianni \etal (2005) to transform the models' $V$ and $I$ photometry
into F606W and F814W magnitudes, We found that the Besan\c{c}on model
and the TRILEGAL model both produced $\approx$ 37 (25) Galactic
foreground stars within our ACS (WFC3) field, down the adopted 90\%
completeness limit of \mi~$\approx$ 27.7, corresponding to 3.2 stars per
arcmin$^2$ over the entire magnitude and color range. The expected
number of foreground Galactic stars increases slightly at our 50\%
completeness limit of \mi~$\approx$ 28.6 (40 and 27 for the ACS and WFC3
fields, respectively). Below these limits, the two Galactic models
diverge substantially, with the Besan\c{c}on model predicting 70\% more
stars at fainter magnitudes. The large offset in predicted counts
between the two Milky Way models at faint magnitudes has been noted by
others (\eg Fantin \etal 2017; Majewski \etal 2017), and is due to
differing model assumptions in both stellar evolution and in the
structural parameters of the Milky Way. However, these results show that
except for brighter regions of our CMDs that contain a small number of
objects, the effect of the foreground contamination from Milky Way stars
is negligible. Instead, as has been found by many others (\eg Ferguson
\etal 2000; Ellis \& Bland-Hawthorn 2007; Windhorst \etal 2011), for deep
HST imaging the dominant source of contamination is unresolved
background galaxies. Although this contamination can be
calculated approximately (\eg Ellis \& Bland-Hawthorn 2007), for an
independent assessment of background contamination and to act as a
background field for the WFC3 pointing, we measure this contamination
empirically from archival HST imaging data as described below.

Because of the extreme depth of our imaging, most of the background
fields used in other studies of the outskirts of nearby galaxies (\eg
Radburn-Smith \etal 2011, Monachesi \etal 2016) have insufficient depth
for our purposes. For our study, we need data that 1) has been taken
using the ACS camera, 2) is equally deep as our observations in both
\mv\ and \mi, and 3) is not known to have a large number of foreground
sources. Ideally, given the known degradation of the ACS detectors with
time (\eg Anderson \& Bedin 2010, Massey \etal 2014, Bohlin 2016, and
references therein), it would also be beneficial to also use background
fields taken close in time to our observations. Only a handful of
datasets satisfy these criteria, from which we chose the Hubble
Frontier Field parallel fields (Lotz \etal 2017) from the Abell~2744
imaging (GO 13495; PI J. Lotz; GO 13386; PI S. Rodney, GO 13389; PI B.
Siana) to estimate the background contamination in our study of M101.

We note several possible concerns with using the Abell~2744 parallel
field for background estimation. First, projected only $\approx$
6.0\arcmin\ (2 Mpc) from the cluster core, the field may contain an
excess of sources from the cluster itself. However, Lee \& Jang (2016)
compared the luminosity functions of objects in the parallel field to
that in both the main Abell~2744 cluster field and the HUDF/XDF field
(Illingworth \etal 2013) and found only a small excess of resolved
galaxies in the Abell~2744 parallel field. Second, due to enhanced
magnification from weak gravitational lensing, the apparent fluxes of
the background sources may be enhanced. Castellano \etal (2016) and
Lotz \etal (2017)  discuss the weak-lensing magnification for the Abell~2744
parallel field, and report a median magnification factor of $\approx$
1.2, depending on the exact position and redshift of the background
source. This magnification bias affect our contamination estimate by
boosting otherwise undetected sources above our detection threshold (\eg
Schmidt \etal 2009) and may lead to a slight overestimate of the
contamination in the M101 field. Third, by basing our background
estimate on a single deep field with small field of view, our estimate
is likely to suffer from the effects of significant cosmic variance (\eg
Somerville \etal 2004; Trenti \& Stiavelli 2008).

With these caveats in mind, we extracted a subset of the total Abell
2744 parallel observations to best mimic the depth of our own science
observations: we used all long F606W visits (eight in total, with five
visits taken from the Frontier Fields, and three visits taken by other
observers), and the first seven visits in F814W from the Frontier
Fields. The F814W exposures consisted of two orbit visits with an
exposure series of (1125s, 1207s, 1307s, 1207s) or (1225s, 1307s, 1507s,
1207s), with the corresponding F606W exposure series being identical in
length. Both Frontier Fields ACS exposure series were effectively
dithered by the primary WFC3/IR pattern ``IR-DITHER-BLOB" and there was
a small random offset between each visit (see Lotz \etal 2017 for full
details). The additional F606W visits were observed at nearly the same
roll angle as the Frontier Fields, and consisted of four 1250s exposures
using the ``WFC3-UVIS-DITHER-BOX" pattern. The total exposure times for
our background field are 35,522s in F814W and 40,430s in F606W. While
additional F814W images are available to increase the photometric depth
in F814W, because our goal is to mimic as closely as possible a
background field that has similar photometric depth and signal-to-noise
as our own M101 data, we did not include them. Given that we are limited
by the availability of F606W background images, and that color
information is critical for our analysis, it is unlikely that that the
additional F814W exposures would improve the background determination
substantially.

To extract the background estimate, we used identical procedure image
alignment, drizzling, object detection and photometry as used for our
M101 ACS and WFC3 fields discussed in Section~2.2. We again used DOLPHOT
for object detection and photometry on all images simultaneously, based
a deep image stack derived from multidrizzling all 28 individual F814W
images together. The final area for the stacked background image was
slightly larger than our M101 ACS field, at $\approx$ 11.95
arcmin$^{2}$. We also determined the photometric errors and completeness
limits for the background field by adding and remeasuring 50,000
artificial stars in an identical manner to the M101 target fields,
deriving 90\% (50\%) completeness limit of 27.8 (29.1) in F606W and 26.5
(28.3) in F814W for the background field,. We then applied the identical
selection criteria to the DOLPHOT output for the background field
discussed in \S 2.3 (object type, S/N values, number of frames detected,
CROWD, CHI, and sharpness) to obtain the final CMD. We detected a total
of 3,613 point sources that met our selection criteria, with 197 (1,383)
sources brighter than the 90\% (50\%) F814W completeness limit.

\begin{figure}[]
\centerline{\includegraphics[width=3.5truein]{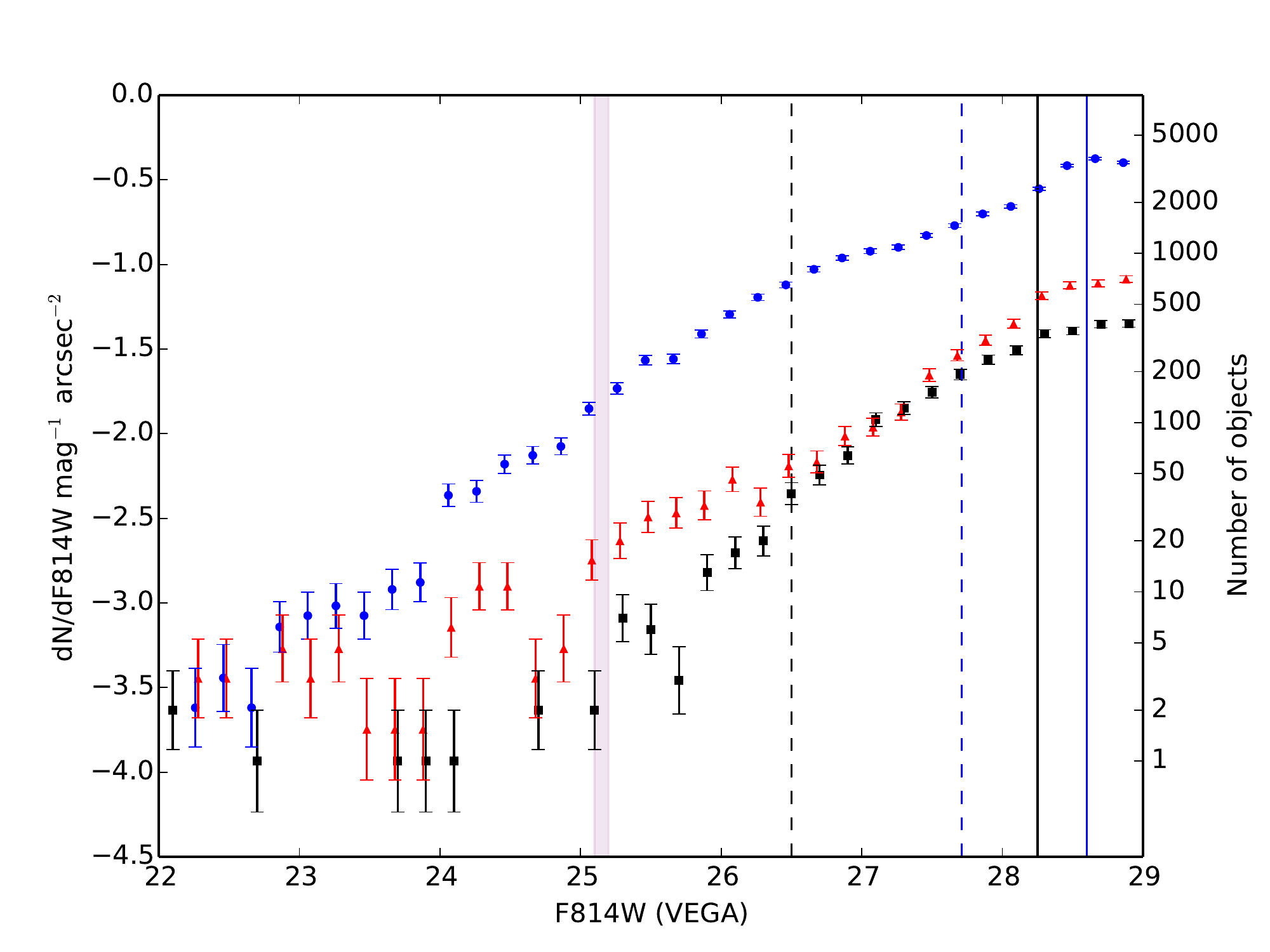}}
\caption{The observed F814W luminosity function of point sources in the
M101 ACS field (blue circles), WFC3 field (red triangles), and background
Abell~2744 parallel field (black squares), binned into 0.2 magnitude
intervals. The shaded purple region shows the approximate location of
the TRGB for M101. The dashed (solid) black and blue lines correspond to
the 90\% (50\%) completeness limits for the background and ACS fields
respectively; the WFC3 completeness limits are within 0.06 magnitudes of
the ACS fields. For ease of comparison, a small offset of $-$0.02
magnitudes and $-$0.04 magnitudes have been applied to the WFC3 and the
ACS number counts bin centers respectively.}
\label{backgroundcomp}
\end{figure}

Figure~\ref{backgroundcomp} shows the F814W luminosity function of all
stellar sources in the background field that passed our selection
criteria, comparing it to the corresponding F814W luminosity functions
from the ACS and WFC3 fields targeting M101. There is a clear excess of
sources in both the ACS and WFC3 M101 fields compared to the background
field. We also note a dip in the luminosity function in both fields just
brightward of \mi~$\approx$~25.2, the approximate location of the tip of
the red giant branch (TRGB) as seen in the CMDs of
Figure~\ref{photcrit}. While the presence of contaminants and AGB stars
at brighter magnitudes makes this ``TRGB edge'' weaker than might be
expected for a pure, extremely old stellar population, one magnitude
{\sl brighter} than the TRGB we find only 338 (26) sources after
background subtraction in the ACS (WFC3) field, while one magnitude {\sl
fainter} than the TRGB there are 1,311 (71) objects. Therefore, even
without using any color information,we have a detection of the TRGB in
both of our observed M101 fields. However, as the luminosity function
extends to fainter magnitudes, the number of background sources and the
systematic uncertainties discussed above increase significantly, and the
color distribution of the sources must be taken into account.

\section{Final CMDs and Stellar Populations}

\begin{figure*}[]
\centerline{\includegraphics[width=7.5truein]{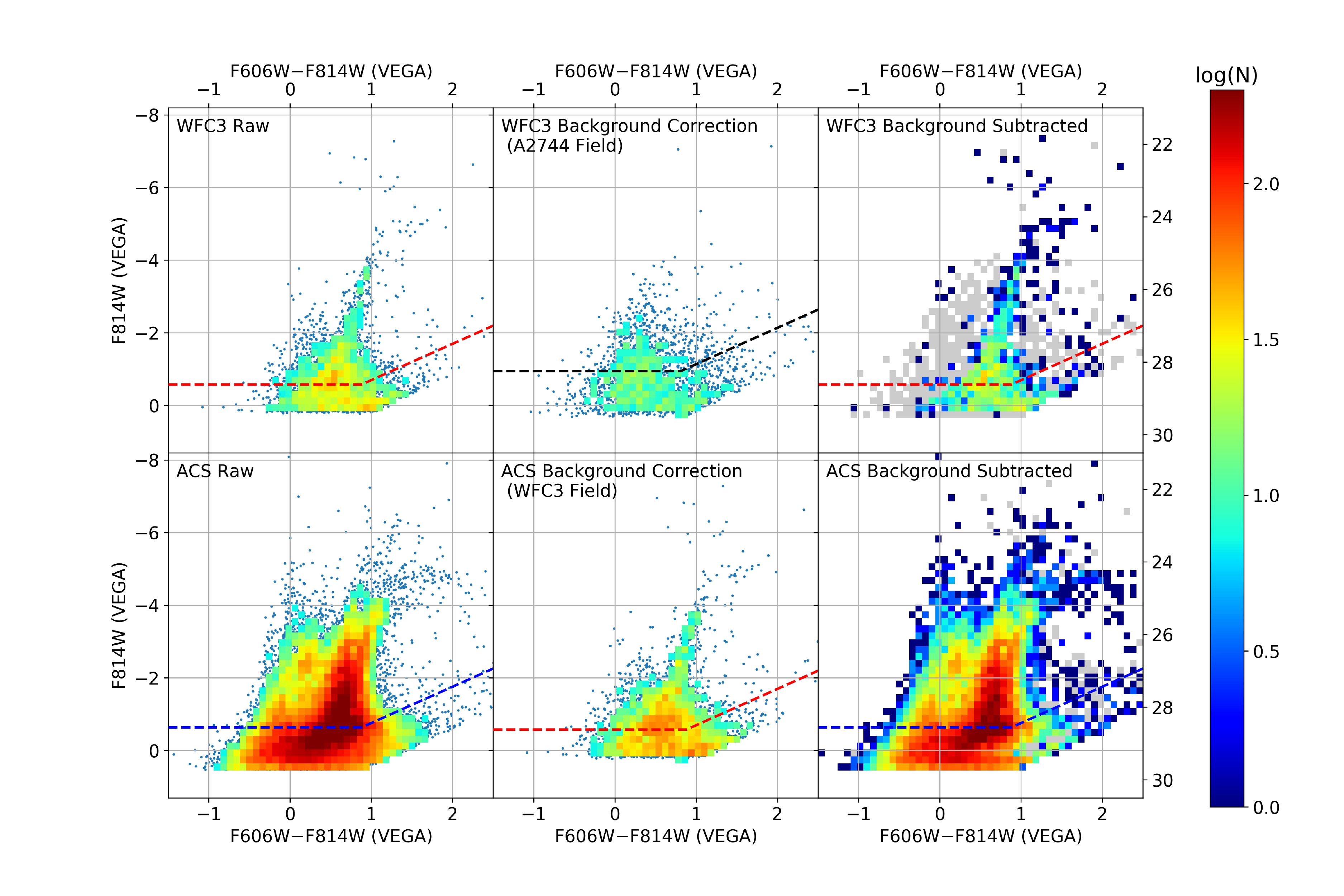}}
\caption{Extracted CMDs for the WFC3 (top) and ACS (bottom) fields. Left
panels show the full CMDs; middle panels show the expected background
contamination estimated using data from the Abell 2744 Flanking Field
(for the WFC3 field), or the WFC3 field (for the ACS field); right
panels show the background-subtracted CMDs. The colorbar shows the
number of stars in the binned regions of the CMDs, and this color scale
is used in all subsequent CMDs shown in this work. The 50\% completeness
limits for each field are shown as dotted lines.
}
\label{CMDback}
\end{figure*}

Figure~\ref{CMDback} shows the cleaned CMDs for the WFC3 and ACS fields,
both before and after background subtraction, and after applying
foreground extinction corrections of $A_{F606W}=0.024, A_{F814W}=0.015$
from Schlafly \& Finkbeiner (2011). For the WFC3 field, the background
correction is taken to be the A2744 CMD, downsampled in size to account
for the different areas of the two fields. For the ACS field, we have
two choices for a background: the A2744 field, or the WFC3 field itself.
In the ACS field, potential contamination comes from two major sources:
unresolved background galaxies, and stars in M101's stellar halo.
Brighter than \mi~=~27, these M101 halo stars dominate over background
galaxies, and have the potential to introduce a false RGB signal into
the ACS CMD. Thus, for the ACS field, we adopt our parallel WFC3 CMD as
the background estimate, with the important caveat that M101 halo stars
could be even more prevalent in the ACS CMD, due to its smaller
galactocentric radius; we examine this possibility in greater detail in
Section 4.1 below. Because the WFC3 field has a smaller area than the
ACS field, we upsample the WFC3 sources by randomly duplicating sources
and adding a random scatter to their \mv\ and \mi\ magnitudes taken from
a uniform deviate spanning $\pm$0.05 magnitudes. For each of the ACS and
WFC3 fields, after construction of the appropriate background estimate,
the CMD and its background are both binned into color-magnitude bins of
size $0.08\times0.2$ mags, after which the background is subtracted off.

In the WFC3 field, the obvious RGB sequence visible in the raw CMD
remains intact after subtraction, along with some hint of an
intermediate age AGB population extending to brighter and redder colors.
Aside from that, we see no clear signature of any other stellar
populations; the scattering of stars blueward of \vmi~=~0.5 and
\mi~$>$~26 appears consistent with background contamination, as
evidenced in the Abell 2744 flanking field. Indeed, blueward of the RGB,
in a region spanning the color range 0.0~$<$~\vmi~$<$~0.45 and magnitude
$-2.5 <$~\Mi~$< -1.5$, counts in the Abell 2744 flanking field are
actually a factor of two {\sl higher} than the counts in the WFC3 M101
field (the effect of this can be seen in the background-subtracted WFC3
CMD, where gray regions indicate oversubtraction). However, this region
of the WFC3 is sparsely populated; along the RGB sequence, the
oversubtraction is less significant, due both to the increased counts on
the RGB but also because of fewer sources in the background estimate on
this portion of the CMD.

\begin{figure*}[]
\centerline{\includegraphics[width=7.5truein]{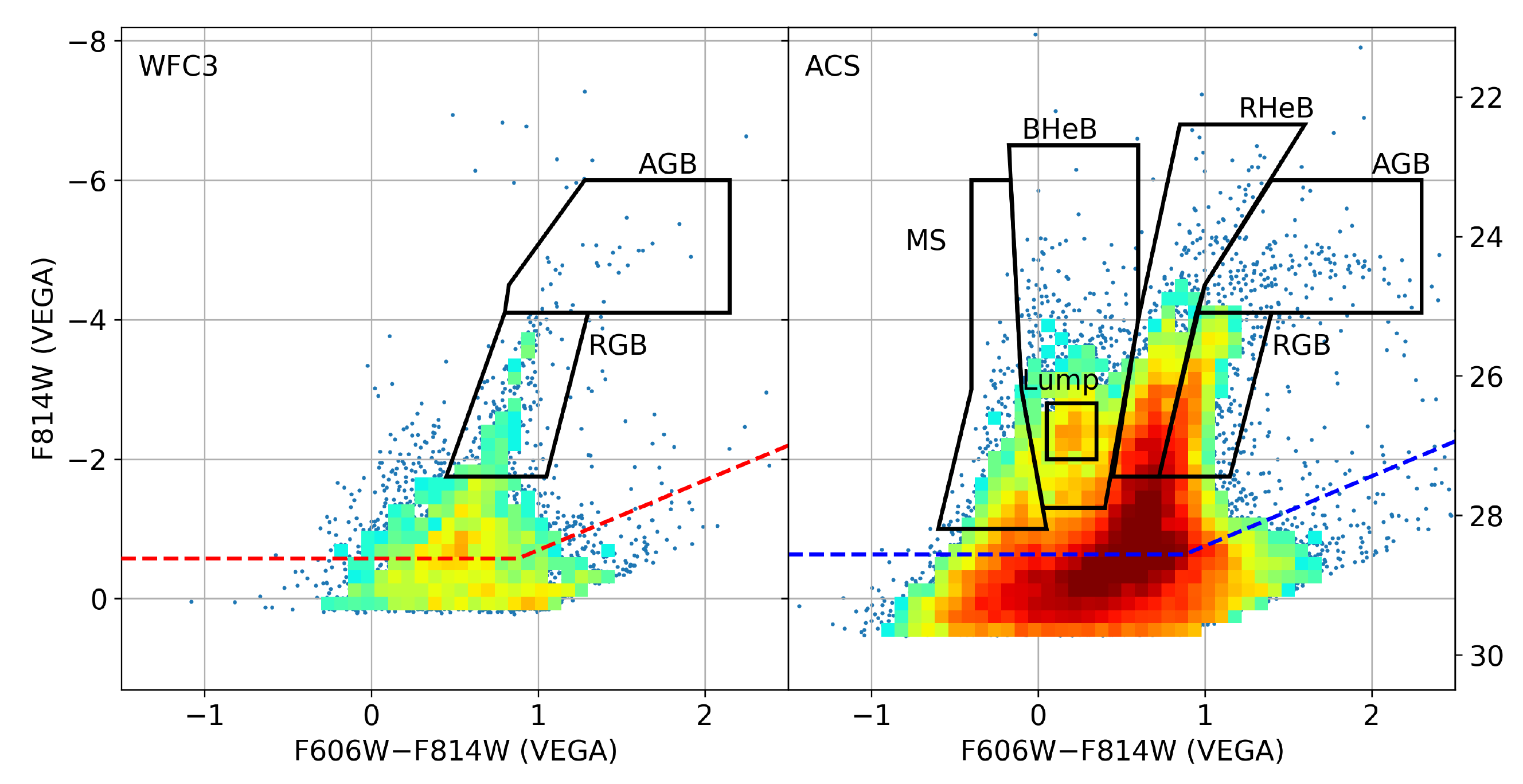}}
\caption{The observed CMDs for the WFC3 and ACS fields. The
  labeled regions show the various stellar populations discussed
  in the text. These boxes also define the photometric selection
  criteria used when performing the quantitative CMD analyses described
  in Section 4.
}
\label{CMDann}
\end{figure*}

In the ACS field, after background subtraction the CMD is largely
unchanged from a qualitative perspective. The multiple stellar
populations evident in the raw CMD remain, and are marked in
Figure~\ref{CMDann}. In addition to the RGB, we see a well-populated AGB
sequence, red and blue helium burning populations (RHeB and BHeB), and
the upper main sequence at \Mi~$\sim-1.5$. The CMD appears qualitatively
similar to those of metal-poor, star-forming dwarf galaxies (\eg McQuinn
\etal 2010), with one notable exception. In star-forming galaxies, the
BHeB typically appears as a smooth sequence of stars extending between
the RHeB and upper main sequence, but in the M101 ACS field the BHeB
sequence instead shows a strong concentration of stars at \vmi~$=0.2$ and
\Mi~$=-2.4$, suggestive of a cohort of stars evolving together with only
a small spread in stellar age and metallicity.

In what follows, we use a combination of isochrone matching and CMD
modeling to further explore the stellar populations in M101's halo and 
outer disk. Having demonstrated the effects of background correction 
in Figure~\ref{CMDback}, our subsequent analyses will focus on the 
{\sl uncorrected} raw CMDs so as to preserve as much information as
possible in the data; in cases where background corrections are necessary
for the analysis (such as for CMD modeling), we incorporate these corrections 
directly into the models, rather than working with the subtracted datasets.

\subsection{Old Stellar Populations in the ACS Disk and WFC3 Halo
Fields}

\begin{figure}[]
\centerline{\includegraphics[width=3.5truein]{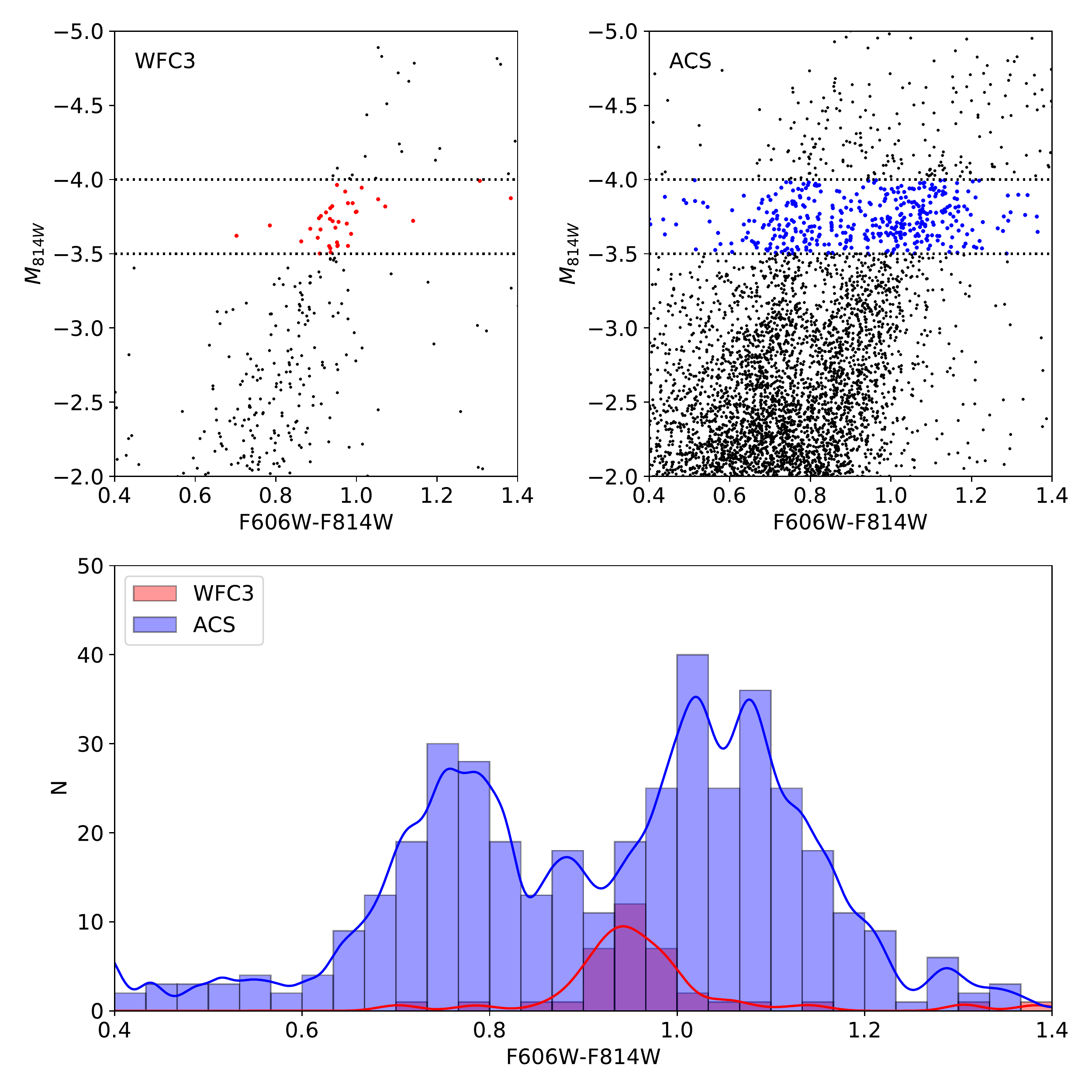}}
\caption{The \vmi color distribution for stars within 0.5 mag of the tip of
the RGB in the WFC3 and ACS fields. The selection is shown in the upper
left (WFC3) and upper right (ACS) panels, while the bottom panel shows
both a binned color histogram and a smoothed representation of the
color distribution (see text for details).}
\label{RGBcomp}
\end{figure}

We begin with a comparative analysis of the old stellar populations seen
in the WFC3 and ACS fields. Both fields show a clear RGB sequence, with
additional AGB stars evident above the RGB tip at \Mi~$\sim-4$
(Figure~\ref{CMDann}). However, due to the paucity of AGB stars in each
field, as well as uncertainties in stellar modeling of the AGB phase, we
confine our quantitative analysis to the RGB sequences, and defer a more
general consideration of the AGB populations to Section 5. The RGB
populations in the two fields are clearly distinct in color from one
another, as can be seen in Figure~\ref{RGBcomp}, which shows the color
distribution of stars within 0.5 mag of the TRGB. We show both a raw
histogram of color, and a smooth distribution function of color calculated as
the sum of gaussians with the means and dispersions set to be the
measured color and color uncertainty of each star in the selection.

In the WFC3 field, the color distribution shows a single peak due to its
old, metal-poor RGB, with median color \vmi~$=0.96 \pm 0.04$. In the ACS
field it shows two peaks; one at \vmi~$\sim$~0.75 due to the RHeB stars,
and another at \vmi~$\sim$~1.05 due to the RGB. For RGB stars within the
color range $0.95 <$~\vmi~$< 1.2$, the median RGB color is \vmi~$=1.07
\pm 0.07$, or about 0.1 mag redder than that in the WFC3
field\footnote{While our use of the Jang \& Lee (2015) photometric
transformation of WFC3 photometry to the ACS system has shifted the WFC3
RGB bluer by $\approx$ 0.04 mag, this is much less than the 0.11 mag
difference we observe in the color of the RGB sequences in the two
fields.}. The RGB population in the ACS field also shows a broader color
spread, suggesting both a higher mean metallicity and a wider range of
metallicities.

Indeed, the broad color distribution of the ACS RGB may be a result of a
superposition of stars in M101's outer disk with those in its halo. The
ACS field is closer to the center of the galaxy than the WFC3 field,
such that any halo contribution could be greater in number or --- in the
presence of a halo metallicity gradient --- redder in color (or both)
compared to that shown in the WFC3 CMD. To assess the possible
contribution of halo stars to the RGB in the ACS field, we examine three
different scalings of the WFC3 populations. The first, most
conservative, model simply takes the WFC3 CMD {\it as observed}, scaling
up in number only to account for the different areas of the WFC3 and ACS
fields. The second model shifts the WFC3 RGB in color redward by 0.11
mags to match the mean color of the ACS RGB, and also scales it up in
number by a factor of 2.27, as might be expected if the halo surface
density profile follows an $r^{-3}$ form (consistent with the outer
halos of MW, M31 and other spiral galaxies; \eg Ibata \etal 2014,
Harmsen \etal 2017, Medina \etal 2018 and references within). The third
model keeps the $r^{-3}$ density scaling, but shifts the color redward
by only 0.075 mags to match the blue side of the ACS RGB color
distribution. We note that either of the color shifts implies a strong
metallicity gradient in the M101 halo. For example, for old (8--10 Gyr)
stellar populations in the PARSEC 1.2S stellar isochrones (Bressan \etal
2012, Marigo \etal 2017; see below for details), a color shift of 0.11
(0.075) mag over 11 kpc would imply metallicity gradients of $\sim
-0.035$ ($-0.02$) dex~kpc$^{-1}$, larger than observed in spiral galaxy
halos to date (\eg Gilbert \etal 2014, Monachesi \etal 2016, Harmsen
\etal 2017).

\begin{figure}[]
\centerline{\includegraphics[width=3.5truein]{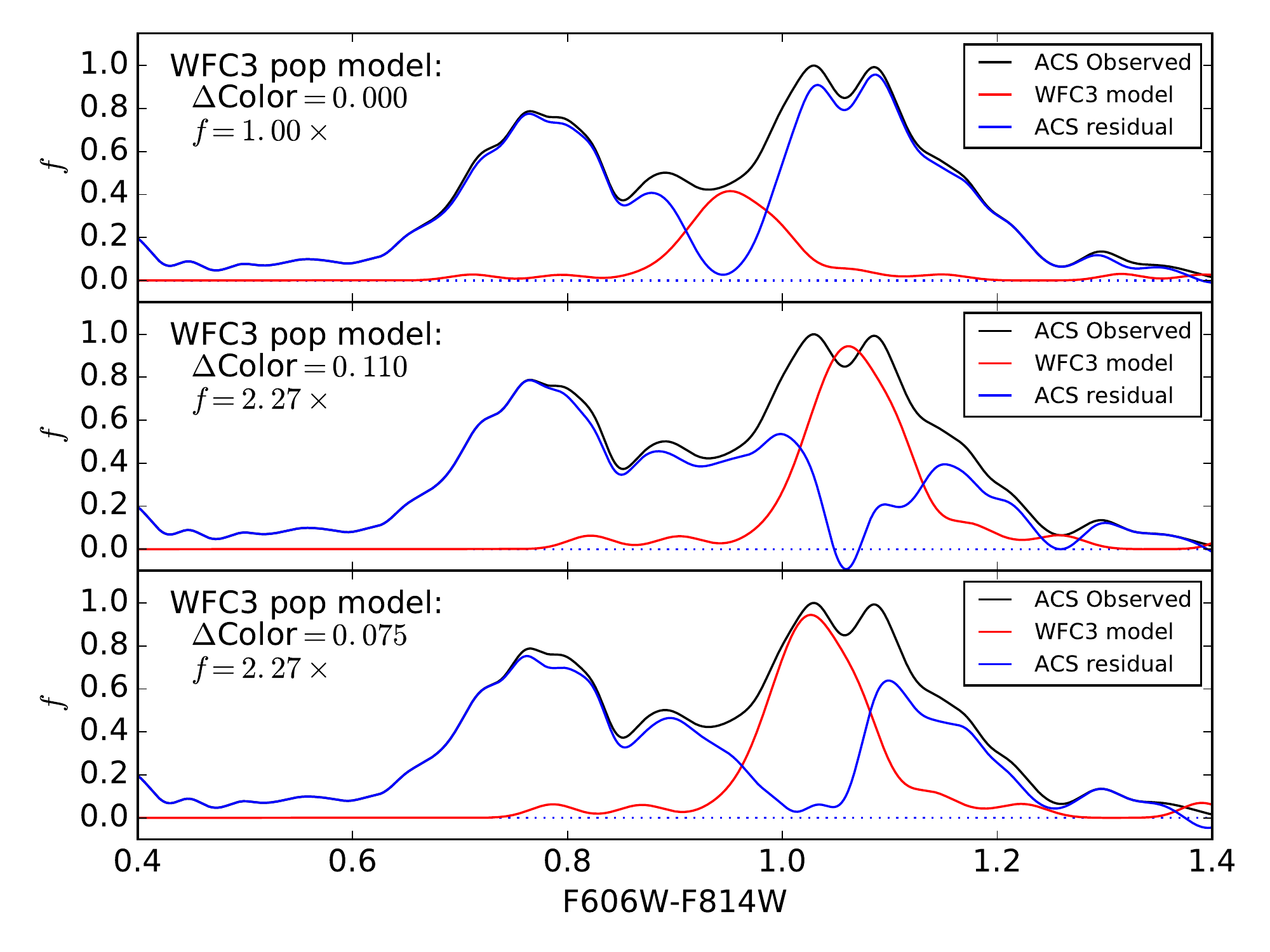}}
\caption{The effect of different halo population models on the color distribution of stars 
within 0.5 of the TRGB in the ACS disk field. See text for details.}
\label{RGBmodel}
\end{figure}

Figure~\ref{RGBmodel} shows the effect of subtracting these different
halo RGB models from the observed color distribution of stars within 0.5
mag of the TRGB in the WFC3 field. Without any color shift or density
scaling, the most conservative model would explain a large fraction of
stars with colors intermediate between the RHeB stars and RGB stars as
coming from the halo population, but not significantly change the
inferred color distribution of the ACS RGB itself. The model with the
strongest color shift, scaled up by the $r^{-3}$ density scaling,
provides a poor match to the observed ACS RGB due to its narrower color
spread --- the model matches the peak number of stars in the ACS RGB but
cannot explain the broad spread in colors. The more modest color shift
of the third model successfully matches the ``blue wing" of the ACS RGB,
and would imply that about 2/3 of the observed ACS RGB is due to a halo
population seen in projection. Again, however, even this model implies a
large metallicity gradient, and is thus likely to be an overestimate of
the halo contamination. A more modest metallicity gradient would push
any halo RGB contamination bluewards, and also force a concurrent
reduction in the number of halo RGB stars so as not to overpredict the
observed counts at these bluer colors. In summary, then, the RGB
population seen in the ACS disk field appears to be only modestly
contaminated by halo RGB stars, and must represent an {\it in-situ}
population of old stars in M101's outer stellar disk distinct in
metallicity from those in M101's stellar halo.

\begin{figure*}[]
\centerline{\includegraphics[width=6.5truein]{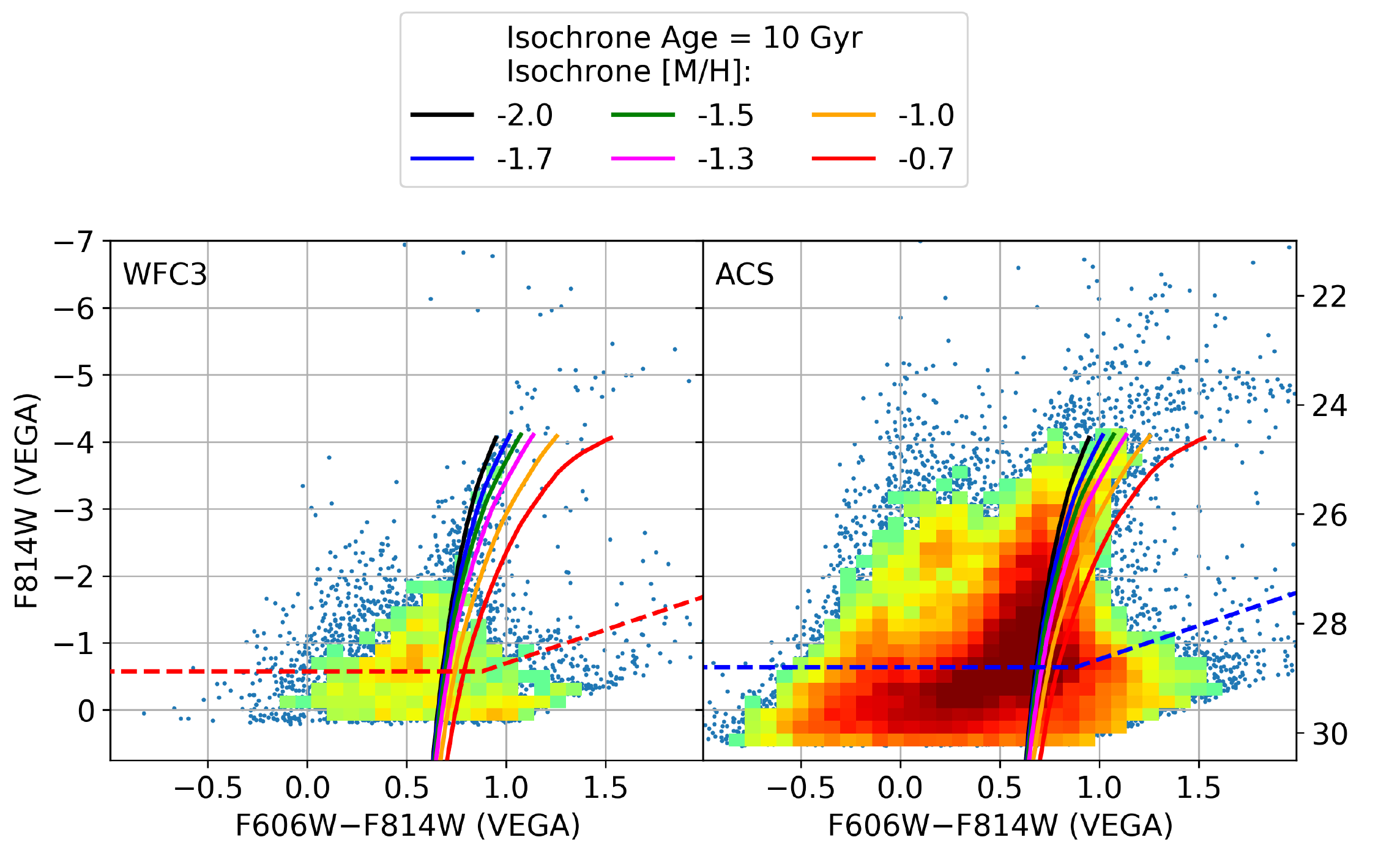}}
\caption{The WFC3 and ACS CMDs with 10 Gyr old PARSEC 1.2S stellar isochrones (Marigo \etal 2017) 
overlaid.}
\label{oldisochs}
\end{figure*}

\begin{figure}[]
\centerline{\includegraphics[width=3.5truein]{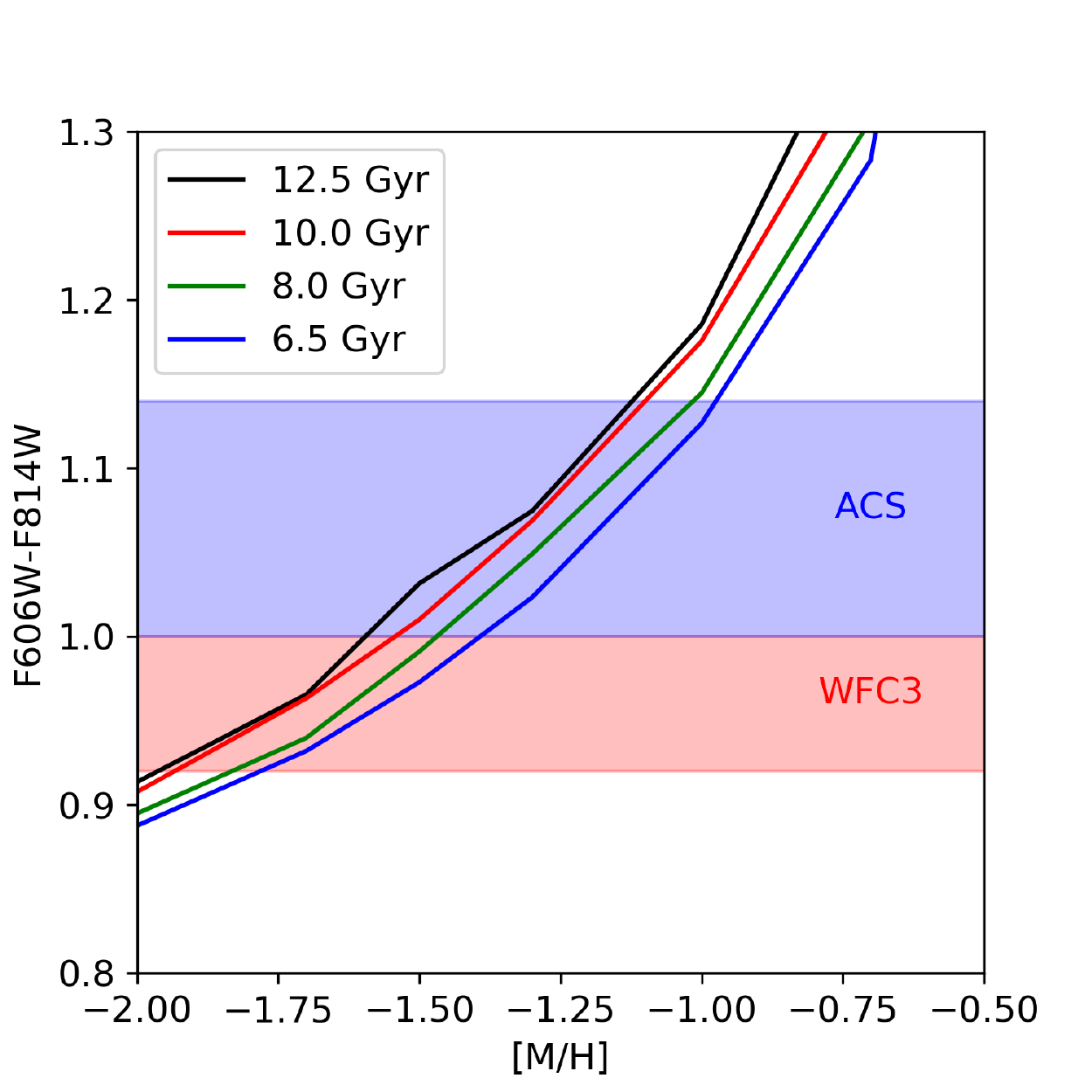}}
\caption{RGB color as a function of metallicity for PARSEC 1.2S isochrones (Marigo \etal 2017)
with ages 6.5, 8, 10, and 12.5 Gyr. The red band shows the $\pm 1\sigma$ color of the WFC3 
RGB, while the blue band shows
the color of the ACS RGB.}
\label{rgbcolor}
\end{figure}

To gain a more quantitative understanding of the stellar populations in
M101's outer regions, we compare the observed CMDs in the two fields to
PARSEC 1.2S stellar
isochrones\footnote{http://stev.oapd.inaf.it/cgi-bin/cmd} (Bressan \etal
2012, Marigo \etal 2017) of varying age and metallicity. These
isochrones use stellar models that presume scaled solar abundances and
adopt $Z_{\sun} = 0.0152$; in this paper we refer to the model
metallicities using the nomenclature [M/H]~$=\log{(Z/Z_{\sun})}$. In
Figure~\ref{oldisochs} we concentrate on the RGB in the two fields,
overlaying 10 Gyr old isochrones with metallicities spanning the range
[M/H]~$=-2.0$ to $-0.7$. In the WFC3 field, the RGB is reasonably well
bracketed by the [M/H]~$=-2.0$ and $-1.5$ isochrones. The RGB in the ACS
field is slightly harder to trace due to the proximity of the RHeB
sequence; fainter than \Mi~$=-2$, the two populations begin to blend
together in the CMD. But at brighter magnitudes, the redder and broader
ACS RGB is bracketed by the [M/H]~$=-1.7$ and $-1.0$ isochrones. We show
this more directly in Figure~\ref{rgbcolor}, which shows the \vmi\ color
of the RGB (again measured as the average color within 0.5 mag of the
TRGB) as a function of metallicity. For the isochrone age of 10 Gyr, the
inferred mean metallicity for the WFC3 RGB is [M/H]~$=-1.7\pm0.2$ and
that for the ACS RGB is [M/H]~$=-1.3\pm0.2$; modestly younger ages shift
the inferred metallicity higher, but by only a small amount~---
approximately 0.2 dex for a 6.5 Gyr old population.

\subsection{WFC3 Field: Surface Brightness}

To better compare our results to deep surface photometry of M101's
integrated stellar light (M+13, van Dokkum \etal 2014, Merritt \etal
2016), we can also calculate the expected surface brightness of the WFC3
field, given the observed RGB and AGB populations in the CMD. We start
with the RGB, by summing the light from all stars detected along the RGB
down to a cutoff magnitude of \mi~=~27.45 (\Mi~=~$-1.75$; see
Figure~\ref{CMDann}). This cutoff magnitude is faint enough to allow us
to probe enough of the RGB without being dominated by the uncertainty
due to background sources and incompleteness corrections at the fainter
magnitudes. To remove background contamination, we selected all objects
within the same RGB selection region in the background Abell 2744
flanking field, and subtract their summed flux from the estimate. To
calculate the total luminosity of the stellar population, we then use
luminosity functions from the PARSEC 1.2S models of Marigo \etal (2017)
to calculate the ``missing light fraction'' ($f_{\rm ML}$) from stars
below our RGB cutoff magnitude. Given uncertainties in the inferred age
and metallicity of the WFC3 RGB, we sample model luminosity functions
that span a range of ages from 6--12 Gyr and metallicity [M/H]~=~$-1.7$
to $-1.5$ and derive a missing light fraction of
$f_{ML,F814W}=0.690\pm0.005$. After correcting for this missing light,
we then add the combined luminosity from {\it observed} AGB stars, as
the PARSEC models do not include any TP-AGB population in their
estimate. To select the AGB population, we adopted the selection region
as shown in Figure~\ref{CMDann}, where we find 24 objects in the WFC3
CMD compared to only 4 in the Abell 2844 background field. We then used
the same methodology for the F606W surface brightness, except using a
subset of the F814W RGB region that included all objects brighter than
F606W~=~27.89 ($M_{F606W}=-1.30$). For this cutoff, the fraction of
missing F606W from fainter stars is $f_{ML,F606W}=0.780\pm0.005$, again
ignoring the AGB contribution.

After applying the model corrections and adding the (observed) AGB
component to our RGB luminosities, we derive a total
$\mu_{F814W}=30.02\pm0.05$ \magsec, where the error is based on bootstrap
resampling of the $F814W$ luminosities for all stars within the RGB and
AGB selection regions, and the uncertainty in the missing light
correction. Similarly, we find, $\mu_{F606W}=30.74\pm0.05$ \magsec\ for our WFC3
field. An important note is these surface brightness results are quite
robust to systematic uncertainties in the background contamination.
Assuming a (generously) large $\pm 30$\% error in our absolute
background subtraction yields an additional shift of $\pm0.05$ in any of
the above results.

To facilitate comparison with previous surface photometry, we convert
the F606W and F814W magnitudes to $V$ and $I$ using the observed
Sirianni \etal (2005) transformations, yielding $\mu_I=29.99\pm0.05$ \magsec\ and
$\mu_V=30.86\pm0.05$ \magsec. These values are fainter than the $\mu_V\sim 29$
mag arcsec$^{-2}$ surface brightness limit of the Mihos \etal (2013)
surface photometry, consistent with the non-detection of any diffuse
light in this particular region. The resulting color of the M101 stellar
population is $V-I = 0.87\pm0.07$ for the entire field, a value
consistent with the colors of metal-poor Milky Way globular clusters
(\eg Harris 1996, 2010), and consistent also with the metallicity of the
RGB stars detected in this field. Finally, to compare our surface
brightnesses to the deep $g$-band surface photometry of van Dokkum \etal
(2014), we convert our $\mu_{V}$ value using transformations from Jordi
\etal (2005) for metal-poor stars. To make use of their $V$ to $g$
transformation, we first adopt a color $B-V =0.65\pm 0.05$, derived
using the $B-V, V-I$ colors of Milky Way globular clusters (\eg Harris
2010). We thus derive $\mu_g=31.10\pm 0.07$ \magsec, where the error is solely
from the uncertainties in our derived surface brightnesses, and not from
any color transformation errors. This derived surface brightness is
$\sim$ 1--2 mag arcsec$^{-2}$ brighter than the {\sl
azimuthally-averaged} surface brightness profile of M101 at this radius
shown in van Dokkum \etal (2014) and Merritt \etal (2016), a sign
perhaps either of azimuthal substructure in M101's halo or systematic
uncertainties in the broadband surface photometry. We comment further on
the comparison with broadband imaging in the Discussion.

\subsection{Young Stellar Populations in the ACS Disk Field}

The young stellar populations seen in the ACS disk field give us the
opportunity to use the the red and blue helium burning sequences to
probe the age and metallicities of those stars as well (see, \eg McQuinn
\etal 2011). As massive stars evolve back and forth across the CMD on
their ``blue loops,'' these stars pile up in the CMD near the turnaround
spot on their evolutionary tracks. Roughly speaking, the BHeB and RHeB
sequences thus trace out these turnaround points in the track as a
function of stellar mass or age --- the more luminous stars on the
sequences are more massive and younger than stars further down the
sequences. Since the turnaround spots on the tracks are also a strong
function of stellar metallicity, we can use the sequences to probe age
and metallicity of the young populations in the field.

\begin{figure*}[]
\centerline{\includegraphics[width=6.5truein]{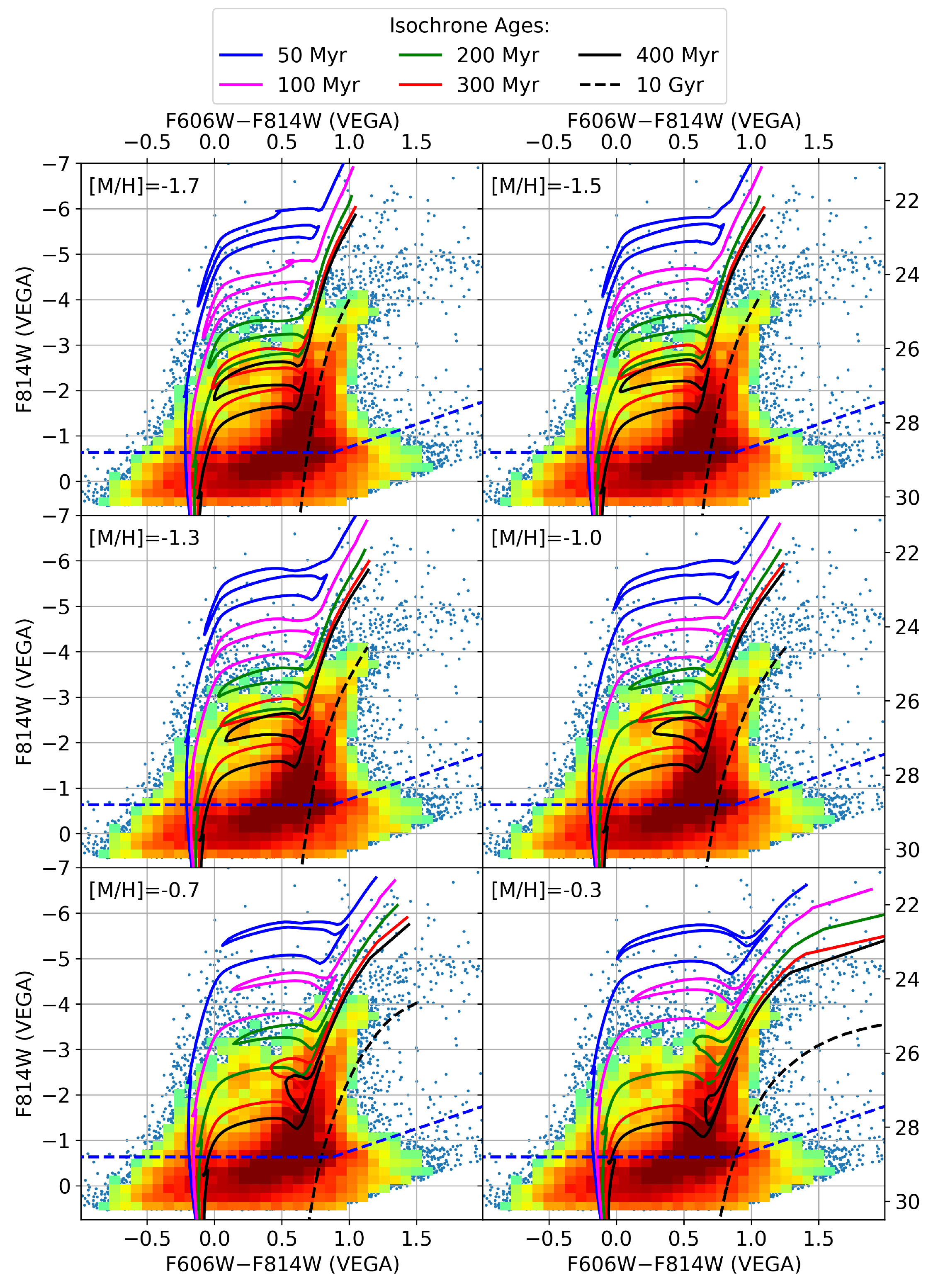}}
\caption{Young stellar isochrones (PARSEC 1.2S; Marigo \etal 2017) 
overlaid on the ACS CMD. Each panel
shows isochrones of varying ages for a given metallicity.}
\label{youngisochs}
\end{figure*}

Figure~\ref{youngisochs} overlays younger PARSEC isochrones for
metallicities in the range [M/H]~=~$-1.7$ to $-0.3$. Of particular
interest here is is the match to blue and red helium burning sequences.
At the lowest metallicities shown ([M/H]~$\leq-1.7$), the model
isochrones place both the red and blue helium burning tracks too blue to
match the observed sequences well. The helium burning tracks at
magnitudes brighter than \Mi~$=-3$ (stellar ages $<$ 100 Myr) reach
colors \vmi~$<0.0$, while the observed BHeB spans a range of color $0.0
< $~\vmi~$ < 0.25$. Similarly, at these metallicities the helium burning
tracks brighter than \Mi~$=-4$ never become redder than \vmi~=~1.0,
whereas the brightest part of the observed RHeB sequence reaches colors
of \vmi~=~1.3--1.4. Conversely, at the highest metallicities
([M/H]~$\geq-0.7$), the helium burning tracks never extend blue enough
to explain the bulk of the observed BHeB sequence. On the red side of
the metal-rich tracks, the most massive stars (ages $<$ 200 Myr) do fall
on the observed RHeB sequence, but metallicities this high cannot then
match the handful of very luminous stars at \Mi~$<-6$, \vmi~$\sim$~1.0
that are explained in the lower metallicity tracks as the youngest, most
massive RHeB stars. Furthermore, at these high metallicities, the model
tracks for stars $\sim$ 300 --- 500 Myr old would fill in the region
between the observed RHeB and RGB populations, while the CMD itself
shows good separation between those populations.

In contrast, metallicities in the range $-1.3\leq$~[M/H]~$\leq-1.0$
reproduce well most features in the observed CMD. On both the blue and
red sides of the helium burning tracks the turnaround colors track the
observed BHeB and RHeB sequences, while the brightest stars (at
\Mi~$<-6$, \vmi~$\sim$~1.0) are explained by a small number of young,
high mass stars in the field. This last population, a sign of very
recent or ongoing star formation, is also seen as a smattering of stars
still on the main sequence with \Mi~$>-4$ and \vmi~$<-0.1$. Based on the
good qualitative match of these isochrones to the young populations in
the CMD, we adopt a final metallicity estimate of
[M/H]~$=-1.15 \pm 0.15$, where the uncertainty estimate simply reflects
the range of isochrones that reasonably match the data.

The inferred metallicity of the young population provides an interesting
comparison to that of the older RGB population discussed in Section 3.
While the young isochrones shown in Figure~\ref{youngisochs} permit
metallicities as high as [M/H]~$=-1.0$, such a metallicity makes for a
poor match to the RGB sequence. The mean metallicity inferred for the
old RGB population in the ACS field is lower by $\sim$ 0.15 dex compared
to the young populations, suggestive that we may be seeing evidence for
chemical evolution in M101's outer disk. However, this metallicity
difference is small, and likely at the limit of our ability to
discriminate metallicity differences given possible age spreads in the
old population as well as uncertainties in the isochrones and stellar
population models.

The helium burning sequences also deliver information on the detailed
history of recent star formation in the field. Along the BHeB sequence,
the distribution of stellar luminosity is a function of the star
formation history, IMF, and evolutionary rate of stars along the helium
burning tracks. For constant (or slowly varying) star formation rates,
the number of stars should drop smoothly at increasing luminosity along
the sequence, simply reflecting the declining numbers of stars at high
mass and their faster rates of evolution. Such behavior can be seen in a
variety of CMDs of star forming dwarf galaxies in the local group
(McQuinn \etal 2011). However, this is not what we observe in the ACS
field, which shows a discrete ``lump'' of stars\footnote{We refer to
this stellar population as ``lump stars" in keeping with the
astronomical tradition of using names that end in ``-ump" to refer to
piles of stars in CMDs --- \eg red clump or AGB bump stars.} in the BHeB
sequence at \Mi~$\approx-2.4$, \vmi~$\approx 0.22$, rather than a
monotonically decreasing number of stars along the sequence.

This BHeB lump is likely a cohort of stars with common age and
metallicity evolving through the blue loop phase. A comparison with the
isochrones in Figure~\ref{youngisochs} shows that for stars in the
metallicity range $-1.3$ to $-1.0$, the lump corresponds to stellar ages
of 300--400 Myr, suggesting this region of M101's outer disk experienced
a weak burst of star formation $\sim$ 350 Myr ago. This age is similar
to the starburst age inferred in M+13, where the very blue colors of the
integrated starlight, coupled with the lack of significant FUV emission
in the NE Plume, also argued for a weak outer disk starburst of age
250--350 Myr. The slightly younger age inferred from the broad band
colors is likely due to the fact that M+13 focused on the integrated
properties of the NE Plume as a whole, whereas our ACS imaging targets a
low surface brightness portion of the Plume which is preferentially
redder and avoids the Plume's most active star forming environments. It
is not surprising, therefore, that we see populations that are slightly
older than those inferred for the Plume as a whole.

While the isochrones shown in Figure~\ref{youngisochs} show the {\sl
positions} of the various stellar population sequences in the CMD, they
do not give information on the {\sl relative numbers of stars}
populating those sequences. As a result, they contain little information
on the age distribution of the young population. In contrast, star
counts along the BHeB sequence --- and in particular the compactness of
the BHeB lump --- are related to the age spread in the evolving cohort
of young stars, with a narrow age spread leading to a more compact lump.
To constrain the duration of the outer disk starburst, we use the
synthetic CMD modeling package IACstar (Aparicio \& Gallart 2004) to
model the effect of the burst duration on the structure of the observed
BHeB lump. As a secondary goal, the IACstar modeling gives a consistency
check on the ages and metallicities inferred from the isochrone matching
shown in Figures~\ref{oldisochs} and \ref{youngisochs}.

To construct a synthetic CMD for comparison to observations, the IACstar
package uses an input star formation history, metallicity model, and
stellar IMF to populate the CMD, for a given choice of stellar
evolutionary library. We again characterize metallicity as
[M/H]~$=\log{(Z/Z_{\sun})}$, although we note that IACstar adopts
$Z_{\sun}=0.02$, which introduces a slight 0.1 dex shift in the [M/H]
values when compared to those inferred from the PARSEC isochrones which
adopted $Z_{\sun}=0.0152$. This shift is likely less than the
overall uncertainty in the CMD modeling, and so for consistancy of
comparison we place the IACstar models on the same solar metallicity
zeropoint as the PARSEC models by setting $Z_{\sun}=0.0152$ when
converting metallicity from $Z$ to [M/H]. Finally, when modeling the
CMDs, we adopt a Kroupa (2001) IMF along with the Teramo stellar
libraries (Pietrinferni \etal 2004).

We start by generating model CMDs for a variety of burst populations,
modeling the star formation histories as Gaussian in time. These
Gaussian bursts peak at a time $t_{\rm burst}$ in the past, and have a
duration given by $\sigma_b$, and a metallicity [M/H]. In modeling
the synthetic CMDs with IACstar, we found the models best match the
location of stars in the lump and RHeB sequence with a metallicity of
[M/H]~$=-1.25$, slightly lower than that inferred from the PARSEC
isochrones ([M/H]~$=-1.15$), even after adjusting for the differing solar metallicity
zeropoints. We thus designate our fiducial model as one with burst age
$t_{\rm burst}=300$ Myr, burst duration $\sigma_b=50$ Myr, and burst
metallicity [M/H]~$=-1.25$, then vary the burst parameters around this
fiducial model to explore the effect of burst age, burst spread, and
metallicity on the structure within the helium burning sequences. To
construct the model CMDs, we begin by using IACstar to generate a CMD
for each model with 100,000 synthesized stars brighter than M$_V =
+1.3$, significantly fainter than our 50\% completeness limit. We turn
the absolute magnitudes into observed apparent magnitudes using our
adopted M101 distance modulus of $(m-M)_0=29.19$, and also add
foreground extinction from Schlafly \& Finkbeiner (2011). We then add
noise to the synthetic CMD using the artificial star tests from
Section~2. In short, for each model star, we match it to an artificial
star closest in color and magnitude, and give it the photometric error
found in that artificial star. In the case where the matched artificial
star was not recovered in the artificial star test, we delete the CMD
star from the synthesized CMD. In this way we match the synthesized CMD
to the observed CMD in both photometric uncertainty and completeness.
Finally, we downsample the total number of stars in the synthesized CMD
so that it has the same number of stars in the region of the BHeB lump
as are observed in the real data. For this exercise, we do not add
background contamination, since we are interested in the {\it intrinsic}
structure of the young populations in the CMD.

\begin{figure*}[]
\centerline{\includegraphics[width=7.5truein]{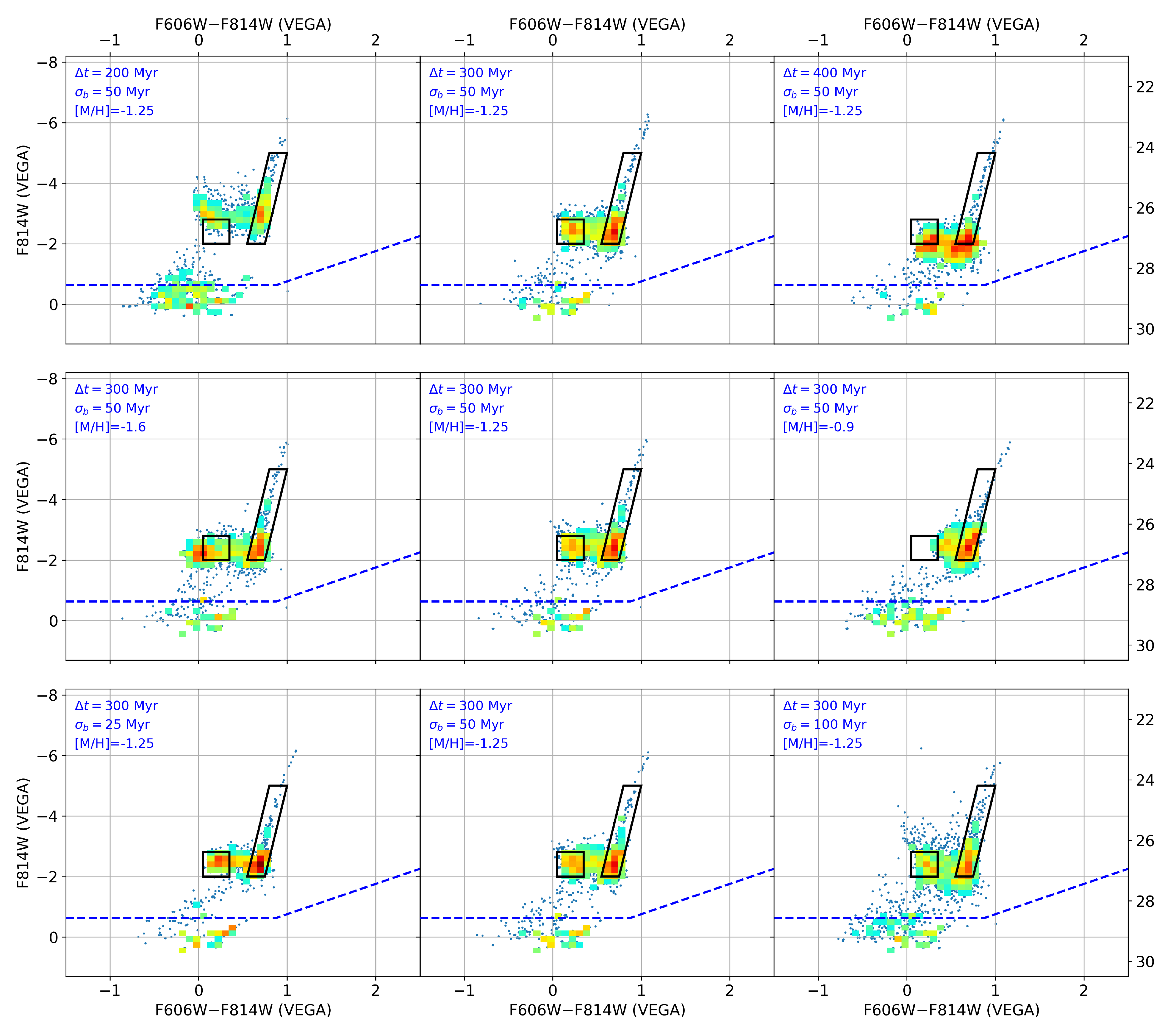}}
\caption{IACstar model CMDs for young starburst populations of different
  metallicity. In each case, the star formation history of the model is
  given by a gaussian which peaked at a time $\Delta t$ in the past and
  had a width $\sigma_b$. The top row varies the burst age $\Delta t$, the middle
  row varies the burst metallicity [M/H], and the bottom row varies the burst
  duration $\sigma_b$. The boxes show the positions of the observed
  BHeB lump and RHeB sequence in the outer disk CMD.}
\label{iacburst}
\end{figure*}

Figure~\ref{iacburst} shows the resulting model CMD for these various
starburst scenarios. In each panel, the location of the BHeB lump in the
observed ACS CMD is shown by the black box, while a parallelogram marks
the spine of the RHeB sequence (with no old stars in these burst models,
there are no RGB or AGB populations). The top row of the figure shows
the results of varying the burst age. It can be clearly seen that
varying $t_{\rm burst}$ by $\pm$100 Myr moves the lump in the synthetic
CMD significantly away from the observed region; more recent bursts show
lumps that are brighter and somewhat bluer than the observed lump, while
older bursts are systematically redder and fainter. The middle row shows
the effect of changing metallicity; at [M/H]~$=-1.6$ the BHeB lump is
too blue, while at [M/H]~$=-0.9$ both the BHeB lump and the RHeB spine
have moved too red. Finally, the bottom row shows changing burst
duration. With a very highly synchronized burst of $\sigma_b=25$ Myr,
the BHeB population shows a very narrow range of absolute magnitude, and
the lump is actually {\it too} concentrated, with no stars along the
broader BHeB sequence. As the burst duration is increased to
$\sigma_b=100$ Myr, stars begin to populate a broader range of the BHeB
sequence, but the discrete lump in the CMD is smeared out, making it a
qualitatively poor match to the observed CMD. These models thus suggest
a reasonable estimate for the burst duration to be $\sigma_b<100$ Myr.

\begin{figure*}[]
\centerline{\includegraphics[width=7.0truein]{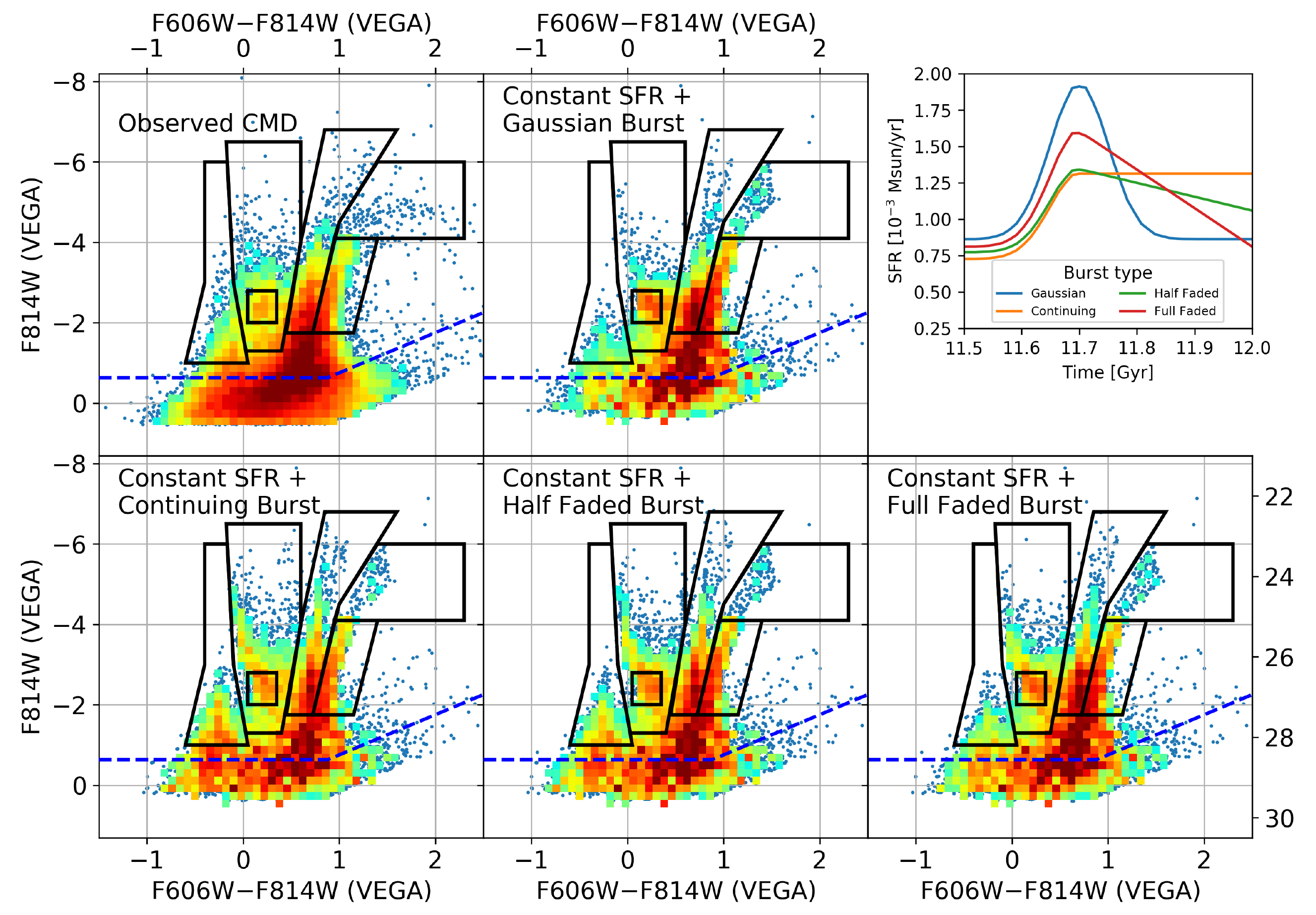}}
\caption{IACstar model CMDs for composite stellar populations
  consisting of a constant star formation history plus recent bursts of
  varying star formation histories. The top row shows the observed CMD
  and the fiducial Gaussian burst model, while the bottom row shows
  three variants of the model where the late time history of the burst
  is varied. The star formation models are shown in the inset panel in the
  top row. All CMDs are modeled using a fixed metallicity of [M/H]~$=-1.25$.
  See text for details.
}
\label{IACcomp}
\end{figure*}

To fully model the observed CMD requires a detailed reconstruction of
the star formation and chemical evolution histories behind the multiple
stellar populations seen in the field --- the older RGB stars, the
post-starburst population seen in the BHeB, and younger stars from
ongoing star formation, as well as the population of halo stars
projected onto the field. Such a reconstruction is beyond the scope of
our intent here; instead, we compare the observed CMD to ones modeled
using a simple ``disk plus burst" star formation history. The disk
population is modeled using a constant star formation rate over the past
12 Gyr and gives rise to the old and intermediate age populations seen
in the CMD. For simplicity, we use a fixed metallicity of [M/H]~$=-1.25$
for the disk population, based on the best matching isochrones shown for
the old population in Figure~\ref{youngisochs}. The fiducial starburst
population is then characterized by a Gaussian with $t_{\rm burst}=300$
Myr, burst duration $\sigma_b=50$ Myr, and metallicity [M/H]~$=-1.25$
based on the best matches shown in Figure~\ref{iacburst}, although we
also run a series of starburst models with varying post-burst behavior
(see below). We add observational noise to the populations as described
previously, and then combine the populations, setting the relative
fraction by mass of the disk and burst populations by matching the
relative numbers of stars observed on the RGB (brighter than
\Mi~=~$-1.75$; fainter than this the RGB and RHeB begin to overlap) to
those found in the BHeB lump ($N_{\rm RGB}/N_{\rm lump}=3.17$). For the
models considered here, the burst fraction $f_b=M_{\rm burst}/M_{\rm
disk}$ ranges from 6.5\% to 14.5\%, depending on the details of the
burst history (see Table~\ref{IACmods}).

\begin{deluxetable*}{lccccc}
\tabletypesize{\scriptsize}
\tablewidth{0pt}
\tablecaption{Composite IACstar Models\label{IACmods}}
\tablehead{
\colhead{ } & \colhead{Observed} & \colhead{Const SFR +} & \colhead{Const SFR +} & \colhead{Const SFR +} & \colhead{Const SFR +}\\
\colhead{ } & \colhead{Properties} & \colhead{Gaussian Burst} & \colhead{Continuing Burst} & \colhead{Half Faded Burst} & \colhead{Full Faded Burst}\\
\colhead{ } & \colhead{ } & \colhead{$f_b=0.065$} & \colhead{$f_b=0.145$} & \colhead{$f_b=0.110$} & \colhead{$f_b=0.095$}
}
\startdata
\multicolumn{6}{l}{Star Counts by Population}\\
Population & $N_{obs}$ & $N/N_{obs}$ & $N/N_{obs}$ & $N/N_{obs}$ & $N/N_{obs}$ \\
\tableline
MS & 898 & 0.30 & 1.27 & 0.93 & 0.46 \\
BHeB & 2212 & 0.78 & 0.97 & 0.91 & 0.91 \\
RHeB & 2812 & 1.14 & 0.99 & 1.03 & 1.15\\
AGB & 159 & 1.35 & 1.06 & 1.16 & 1.26 \\
RGB & 1978 & 1.29 & 1.01 & 1.13 & 1.20 \\
Lump & 592 & 1.34 & 0.99 & 1.10 & 1.22 \\
\tableline
\multicolumn{6}{l}{Integrated Light Properties}\\
$\mu_B$ (\magsec) & 28.1 $\pm$ 0.1 & 28.2 & 27.8 & 27.9 & 28.0 \\
$B-V$ & 0.31 $\pm$ 0.05 & 0.39 & 0.24 & 0.29 & 0.34 \\
\enddata
\tablecomments{Populations are defined using the regions shown in
  Figures~\ref{CMDann} and \ref{IACcomp}. The burst fractions are defined
  as $f_b=M_{\rm burst}/M_{\rm disk}$. The observed integrated surface brightness 
  and color of the field come from the imaging of Mihos \etal (2013).}
\end{deluxetable*}

In addition to these model populations, however, we must add an
appropriate background population as well, accounting both for
unresolved background sources and for the presence of M101 halo stars
projected onto the M101 disk. For this, we adopt the WFC3 halo field 
CMD as the appropriate background, since it includes both true background
contamination plus any contamination from M101's stellar halo. To account for
the different FOV of the ACS and WFC3 cameras, we scale up the WFC3
counts by a factor of 1.55 (the relative area of ACS vs WFC3) by randomly
duplicating WFC3 sources after adding a $\pm0.05$ mag scatter to their
magnitudes. We again note, however, that this approach implicitly assumes that
the halo population found in the WFC3 field is identical in number
density to that which contaminates the ACS field and, as discussed
previously, may well underestimate the true halo population in the
synthesized CMDs.

Figure~\ref{IACcomp} shows both the observed ACS CMD and the modeled
CMDs for these various composite star formation histories, while
Table~\ref{IACmods} details the population number counts (relative to
the observed numbers) and integrated light properties for each modeled
CMD. The fiducial pure Gaussian model is shown top center, and does a
reasonable job of matching the observed CMD, with one notable exception:
it has significantly fewer main sequence stars than observed,
particularly those brighter than \Mi~=~$-2$. Such luminous main sequence
stars must have ages $\lesssim$ 200 Myr (see, \eg
Figure~\ref{youngisochs}), arguing that there must be residual ongoing
star formation above that predicted by the simulated star formation
history. This is also borne out by the broadband color of the integrated
light in the model (shown in Table~\ref{IACmods}), which is somewhat too
red compared to the broadband $B-V$ colors in the field derived from the
deep imaging of M+13.

To examine the effects of late time residual star formation, we look at
three variants on the Gaussian burst model: one where, after ramping up,
the SFR stays constant at the peak level to the present day
(``continuing burst''), and two where the SFR declines linearly to half
peak (``half faded") or zero amplitude (``full faded") at
present.\footnote{This does not mean there is no star formation at
present; all models continue to have ongoing star formation from the
constant SFR ``disk" component; see the SFR histories in
Figure~\ref{IACcomp}. Note also that all models are consistent with the
upper limit on current star formation rate in the ACS field ($\approx
0.002\ {\rm M}_\sun\ {\rm yr}^{-1}$) derived from the H$\alpha$ imaging of Watkins
\etal (2017). } These model star formation histories are shown in upper
right panel of Figure~\ref{IACcomp}, and the resulting CMDs are shown in
the bottom panels. The continuing burst model now somewhat overpredicts
the MS population and yields integrated colors that are slightly too
blue, and the BHeB sequence in the CMD appears much more ``stream-like"
than in the observed CMD. In the half-faded burst model, the BHeB shows
more of the discrete lump seen in the data, as well as more closely
matching both the number counts and broadband colors of the region. In
contrast, the fully faded burst model significantly underpredicts the MS
population, resulting in a poorer match to the data. Clearly, these star
formation histories are simple cartoon models for the detailed history
of star formation in M101's outer disk. Nonetheless, the modeled CMDs
corroborate the inferences from the isochrone matching that the stellar
populations trace a recent $\sim$ 300 Myr old weak burst of star
formation in the outer disk which has partially but not completely faded
by the present day.

\subsection{Spatial Distribution of Stellar Populations}

\begin{figure*}[]
\centerline{\includegraphics[width=7.0truein]{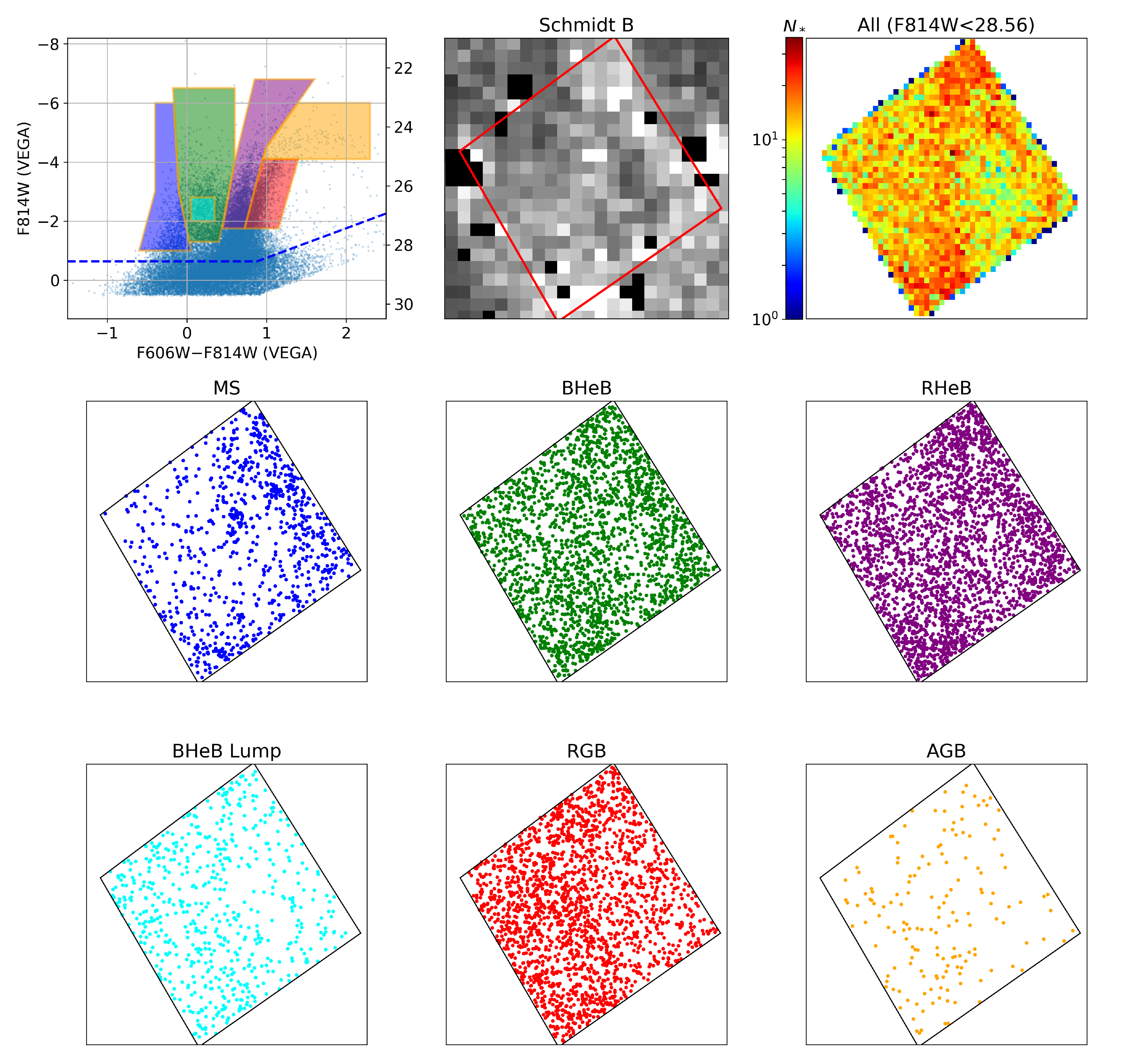}}
\caption{The spatial distribution of stars in the ACS field. The top
  left panel shows the CMD, with colored regions associated with varying
  stellar populations. The top middle panel shows the broadband B image
  from the Burrell Schmidt (M+13) masked and rebinned to show low surface
  brightness structure, while the top right panel shows the spatial
  distribution of all stars brighter than the 50\% completeness limit
  (\mi~=~28.56) in the ACS field. Subsequent panels show the spatial
  distributions of main sequence stars (MS), blue and red helium burning
  stars (BHeB and RHeB), stars in the BHeB lump, and red giant and
  asymptotic giant branch stars (RGB and AGB). }
\label{SpatFig}
\end{figure*}

We can also use the CMDs to trace the spatial distribution of different
stellar populations across the ACS field of the view. While we reserve a
quantitative study of spatial clustering for a future paper, we present
here a broad brush comparison of the spatial distribution of the stellar
populations to the broadband diffuse light. In the top left panel of
Figure~\ref{SpatFig} we show the populations defined on the CMD: main
sequence (MS), blue and red helium burning sequences (BHeB, RHeB), the
co-evol lump stars in the BHeB (lump), and stars on the red giant and
asymptotic giant branches (RGB, AGB). We then plot the location of stars
in each stellar population, along with the deep B-band imaging of M+13,
masked of bright, compact sources and median binned to 13\arcsec\
resolution to show low surface brightness structure. A comparison of all
stellar sources brighter than \mi~=~28.56 (the 50\% completeness limit
of the ACS field) shows good match to the broadband imaging --- the
highest density of stars is found in the north and south corners of the
field and along the NW edge, regions that correspond to a higher
integrated surface brightness of blue starlight. The population of main
sequence stars follows a similar spatial distribution and shows
significant clustering; at these luminosities (\Mi~$<-1$) the main
sequence lifetime is short, less than 100 Myr, such that these stars
have not had time to drift far from their birthplaces. The more evolved
helium burning sequences, which cover a larger age spread up to 500 Myr,
are not as clustered and are weaker tracers of the broadband blue light.
In contrast, the older RGB stars (ages $>$ 1 Gyr) are more smoothly
spread across the field and, if anything, show a slight excess in the
eastern (left) side of the image, unlike the blue light. While the blue
light and massive young stars together follow the spatial distribution
of recent star formation, the RGB stars may be tracing the underlying
mass density of M101's outer disk. Interestingly, sources in the BHeB
lump, which should consist of a co-evolutionary population with age
$\sim$ 300 Myr, seem distributed more like the RGB stars on larger
scales, although we do also see signs of small scale clustering in this
population as well. The spatial disconnect between the lump stars and
the blue light argues that the spatial pattern of star formation across
the field has changed over this 300 Myr timescale, or that stars in the
BHeB lump have drifted far from their original birthsites. The latter
scenario is not unlikely; if young stellar populations have velocity
dispersions of $\sigma_v \sim 10-20$ \kms, similar to that of young
stellar populations in the Milky Way (\eg Dehnen \& Binney 1998), over a
300 Myr span they could have drifted roughly $\sigma_v \times t \approx
3-6$ kpc, comparable in scale to the $\approx$ 7 kpc size of the ACS
field of view.

\section{Discussion}

Our {\sl HST} imaging of fields in the outskirts of M101 provide new
information on the galaxy's star forming history and the properties of
stellar populations in its outer disk and halo. In the ACS disk field,
we see a metal-poor population consisting of both old red giant stars
and young massive stars, and confirm the post-starburst nature of the
populations suggested by the broadband imaging of M+13. In the WFC3 halo
field, we see only older populations --- RGB and AGB stars --- with much
lower metallicities than the disk populations observed in the ACS field.
In this section, we now turn to several open questions remaining: the
metallicity of the outer disk, the origin of the stars in the WFC3
field, the nature of the detected AGB populations, and the properties of
M101's halo compared to those of other galaxies. We then present a the
cross-comparison between the resolved stellar population work presented
here and the inferences from deep broadband imaging of M101's integrated
light. Finally, we frame our results in the overall context of M101's
dynamical history of interactions within the group environment.

\subsection{The metallicity of the outer disk}

The metallicities we measure for the stellar populations in the
outer disk are extremely low ([M/H]~$<-1$), motivating a comparison with
metallicities inferred in the outskirts of other disk galaxies. In doing
this comparison, we note that the IACstar models give a slightly lower
metallicity (by $\sim$ 0.1 dex) than the PARSEC isochrones suggest.
Again, this difference is likely within the uncertainties of the models,
and indeed we note that characterized by metal abundance $Z$, the two
techniques agree; the PARSEC isochrones that bracket the young
populations in metallicity are $Z=0.00076$ and $Z=0.00152$, while the best
match IACstar models use $Z=0.0008$. On balance, we favor the PARSEC
models due to their use of more recent stellar models, and thus adopt a
value of [M/H]~$=-1.15 \pm 0.2$ as our best match metallicity for the
young populations in M101's outer disk, and a slightly more metal poor
[M/H]~$=-1.3 \pm 0.2$ for the older RGB stars. The quoted errorbars
simply reflect the range of isochrone metallicities that bracket the
observed populations shown in Figures~\ref{oldisochs} and
\ref{youngisochs}.

Metallicity gradients in spiral galaxies can be most easily traced
through HII region abundances. These gradients typically follow an
exponential decline with radius, but then often flatten in the disk
outskirts, such that extremely low metallicities are uncommon (see \eg
the review by Bresolin 2017). For example, in M83 there appears to be a
metallicity ``floor'' of $12+\log{O/H} \simeq 8.4$ (Bresolin \etal
2009); for a solar oxygen abundance of $12+\log{O/H} = 8.69$ (Asplund
\etal 2009) and assuming scaled solar abundances, this corresponds to
[M/H]~$\simeq -0.3$. In the Milky Way, both young star clusters and
Cepheid variables also show evidence for a flattened metallicity
gradient beyond 12 kpc, with typical outer disk metallicities of
[M/H]~$\sim -0.3$ (\eg Luck \etal 2003, Yong \etal 2005, Sestito \etal
2008, Pedicelli \etal 2009). Compared to these values, our inferred
metallicity for the young populations ([M/H]~$= -1.15$) is significantly
lower. However, in M101 itself the HII region oxygen abundances show no
clear break from a pure exponential (Kennicutt \etal 2003; Croxall \etal
2016), and the outermost disk HII regions show metallicities of
$12+\log{O/H} = 7.55$ (Kennicutt \etal 2003), equivalent to [M/H]~$=
-1.14$, in quite good agreement with our inferred metallicity for the
young population. This match between the gas-phase metallicity and that
of the stellar populations we see in the field is a strong argument that
these populations were formed largely {\it in-situ} rather than being
scattered or migrating populations from the more metal-rich inner disk.

Metallicities for old stellar populations in disk outskirts are harder
to come by. The low surface brightness of these regions makes all but
the nearest disk galaxies difficult to study via means other than
integrated colors, which suffer from the well-known age-metallicity
degeneracy and thus provide only the crudest of information on stellar
metallicities. Instead, resolved population studies have largely been
confined to a relatively few disk galaxies found within a few Mpc of the
Milky Way. In M31, for example, studies of resolved RGB stars show that
the metallicity gradient in the disk flattens at large radius; beyond 20
kpc typical metallicities are [M/H]~$\sim -0.5$ (\eg, Worthey \etal
2005, Bernard \etal 2015) , significantly higher than what we find for
M101. In contrast, while the lower luminosity Sculptor Group spirals NGC
300 and NGC 7793 also show flat outer disk metallicity profiles, their
outskirts show much lower metallicities of [M/H]~$\simeq -1.0$ and
$-1.5$, respectively (Vlaji\'{c} \etal 2009, 2011). The metallicities we
infer for the old populations in the ACS field ([M/H]~$= -1.3 \pm 0.2$)
are similarly low.

\subsection{The origin of stars in the WFC3 field}

We targeted the WFC3 camera in a blank field away from the observed
stellar disk in order to sample populations in M101's stellar halo.
However, one concern with the interpretation of the WFC3 field as a halo
field is its proximity to M101's distorted outer disk. Projected only a
few kpc away from the extended Northeast Plume, the stars we detect in
the WFC3 field could in principle be an extension of M101's disk to even
larger radius, or a population of stars scattered outwards from M101's
disk due to interactions between M101 and its companions.

However, the mix of stellar populations in the two fields is quite
different, with no sign of a young population in the WFC3 field. If the
populations in the ACS and WFC3 fields were similar, the helium burning
sequences should be visible in the WFC3 CMD, even accounting for its
smaller number of stars. We see no such sequences, and the smattering of
blue stars seen in the CMD is consistent with that expected from
background contamination (see Figure~\ref{CMDback}). Based on the lack
of a young population alone, the WFC3 field is not simply an extension
to lower surface brightness of the star forming populations seen in the
ACS field.

However, the lack of young stars does not preclude contamination from an
extended disk component. Many disk galaxies show red color gradients in
their outskirts, signifying an older stellar population extending beyond
the star forming disk. The color difference between the RGB populations
in the WFC3 and ACS fields (Figure~\ref{RGBcomp}) again argues that
these two populations are not identical, but in principle these
differences are consistent with observed metallicity gradients in galaxy
disks. In the Milky Way, for example, metallicity gradients in young
populations such as Cepheids (Pedicelli \etal 2009, Luck \& Lambert
2011) or open clusters (Friel \etal 2002, Sestito \etal 2008) range from
$-0.05$ to $-0.09$ dex~kpc$^{-1}$, with suggestions of shallower
gradients at large radius (beyond 10 kpc; \eg Yong \etal 2005, Carraro
\etal 2007, Sestito \etal 2008) or when traced by older populations
(Cheng \etal 2012). In M101 itself, gas-phase oxygen abundances in HII
regions have radial gradients of $-0.03$ dex~kpc$^{-1}$ (Kennicutt \etal
2003, Croxall \etal 2016). This gradient is very similar to the
metallicity gradient we would infer by comparing the RGB populations in
the ACS and WFC3 fields (d[M/H]/dr = $-$0.035 dex~kpc$^{-1}$), although
of course the RGB metallicities themselves are down by $\sim$ 0.35 dex
compared the measured gas-phase metallicities.

Based on the stellar populations alone, therefore, it is hard to
discriminate between the WFC3 populations being indicative of a true
halo population versus that of an extended outer disk. Morphologically,
though, the deep surface photometry of M+13 shows that the outer disk is
quite distorted, with no smooth outer component. While azimuthally
averaged the disk shows a quasi-exponential structure, beyond about
14\arcmin\ (28 kpc) there is little evidence for any azimuthal symmetry;
the outer regions of the disk consist solely of the tidal plumes.
Therefore, it is unlikely that the WFC3 field samples a dynamically
relaxed outer disk that is easily characterized by a smooth metallicity
gradient. If this field is sampling disk stars, it is more likely they
would be stars scattered outwards by interactions. In principle, if the
scattering time is longer than the ages of the helium burning stars ($>$
a few hundred Myr), only the RGB and AGB stars would be present in the
CMD, such that the lack of young stars isn't an iron-clad argument
against disk scattering. However, under that scenario, we would expect
metallicities of the RGB stars in the field to reflect those found at
smaller radius, like those in the ACS field or even more metal-rich
stars from the inner disk. Yet the color distributions of the RGB stars
in the WFC3 and ACS fields are demonstrably different (as shown in
Figures~\ref{RGBcomp} and \ref{RGBmodel}). Thus, while we cannot
absolutely rule out a disk scattering origin for these stars, we find it
more plausible that we are sampling stars tracing M101's halo population
instead.

\subsection{AGB Stars in the WFC3 Field}

The CMDs shown in Figure~\ref{CMDann} also clearly show the presence of
asymptotic giant branch (AGB) stars in both the ACS and WFC3 fields -
more specifically, the thermally-pulsing AGB (TP-AGB) stars that are
more luminous than stars at the RGB tip. They are of particular interest
in the WFC3 field where the RGB stars alone provide little information
on the timescale of star formation in the field. The presence of TP-AGB
stars significant brighter than the RGB tip can be an indicator of an
intermediate-age ($t\sim$ 1--10 Gyr) stellar population for metal-poor
([M/H]~$<-1$) stars (\eg Guarneri \etal 1997, Gallart \etal 2005, Grebel
2007). Thus their presence in the WFC3 field provides an age
discriminant (albeit approximate) for the old metal-poor population
present in that field.

We first test the significance of this population through star counts in
the WFC3 CMD. Above the RGB tip at \Mi~=~$-4.1$ lies a population of
objects that extend to \Mi~=~$-5.5$, with colors ranging from that of
the RGB tip to redder colors of \vmi~$\approx$~2.0. In this region of
the CMD we observe $N=22$ objects, ignoring objects immediately around
\Mi~=~$-4.1$ which are likely RGB stars. We can estimate the Milky Way
foreground contamination using the TRILEGAL and Besan\c{c}on models
described in Section 3, and find $3.2\pm1.6$ (TRILEGAL) and $3.5\pm1.8$
(Besan\c{c}on) objects, which compares favorably with the $N=4$ objects
observed in the Abell 2744 background field. We thus find a
background-corrected excess population of $N_{AGB}=18\pm5$ objects
(using Poisson uncertainties only) in the WFC3 CMD, or a $>3\sigma$
detection of an AGB in this field.

While the small number of AGB stars precludes a detailed analysis, we
can compare these objects to AGB populations seen in other resolved
stellar populations to place rough constraints on the star formation
history of the field. One simple metric is to measure the maximum
brightness of the AGB stars above the TRGB ($\Delta_{814}$), which at
fixed metallicity should range brighter for younger stellar populations
(\eg Rejkuba \etal 2006). For example, in a detailed analysis of stellar
populations in a dozen metal-poor galaxies taken from the ANGST sample
(Dalcanton \etal 2009), Girardi \etal (2010) found AGB stars typically
extending $\Delta_{814}=1.0-1.5$ magnitudes above the TRGB. The galaxies
in that sample had metallicities in the range ($-1.7<$[Fe/H]$<-1.1$),
comparable to that we infer from the WFC3 RGB population, and have very
little evidence for recent star formation, with 80-95\% of the stars in
the sample galaxies having formed more than 3 Gyr ago. In our WFC3
field, we measure a similar range of AGB luminosities, with
$\Delta_{814}=1.0$ (or 1.4, if the two stars at \mi~=~23.7 are
included). Given the similar AGB extent to the Girardi \etal sample, the
AGB stars seen in our M101 WFC3 field seem consistent with an
intermediate stellar population age of 3 Gyr or older, with no sign of
more recent star formation.

The only difference we see between the M101 halo field and the ANGST
galaxy subsample is the total number of AGB stars present. Girardi \etal
(2010) find AGB-to-RGB fractions ($N_{\rm AGB}/N_{\rm RGB}$) between
0.02 and 0.05, where $N_{RGB}$ is derived using counts within two
magnitudes of the RGB tip. Measured in the same way, our WFC3 RGB
contains $226\pm15$ objects, or a total (corrected for background) of
$N_{RGB}=188\pm17$. Given the $N_{AGB}=18\pm5$ measured above, we derive
$N_{\rm AGB}/N_{\rm RGB} = 0.096\pm0.028$. This value is larger than
that observed in the Girardi sample at the $2\sigma$ level, and may hint
at differences in stellar populations compared to those ANGST galaxies.
However, given the small number of AGB stars seen in the WFC3 field, we
are hesitant to place too much significance to this result.

\subsection{The properties of M101's halo}

If the WFC3 field is in fact tracing M101 halo stars, what inferences
can be made about the nature of M101's halo in general? While the small
FOV of the WFC3 field precludes us from giving a global view of the
entire M101 halo, we attempt here to place our results in context with
other spiral galaxy halos. The stellar halos of spiral galaxies
(including the Milky Way and M31) have very diverse properties with a
wide range of mean metallicities, number density profiles, and
metallicity gradients (\eg Mouhcine \etal 2005ab, Ibata \etal 2014,
Monachesi \etal 2013, 2016, Merritt \etal 2016, Harmsen \etal 2017)
While there is no clear correlation of these properties with {\sl total}
galaxy mass or luminosity, there is a trend with the {\sl stellar halo}
mass, in the sense that lower-mass stellar halos have lower
metallicities and weaker metallicity gradients (Harmsen \etal 2017).
These trends show broad agreement with models in which the outer halos
of galaxies are shaped by accretion, and where galaxies with more
massive stellar halos (\eg M31) have accreted more massive satellites
which increase the mean metallicity of the resulting halo population
(\eg Font \etal 2006, Cooper \etal 2010, 2013, Deason \etal 2016,
D'Souza \& Bell 2017).

The low metallicity we infer for the WFC3 halo field is significantly
more metal-poor than that typically observed in the outer halos of
spiral galaxies to date. Adopting a typical halo value of
[$\alpha$/Fe]$=+0.3$, our metallicity estimate of
[M/H]~$=-1.7$ translates to [Fe/H]$\sim -1.9$ (Salaris \etal 1993;
Streich \etal 2014), lower by nearly 0.5--1 dex compared to most spiral
galaxy halos; the Milky Way's halo is the only one known to approach
these low metallicities (Harmsen \etal 2017). Some of the difference may
be due to the different radial ranges probed; the Harmsen \etal sample
is measured at 30 kpc, while our WFC3 field is further out at 47 kpc.
Our analysis of the RGB colors in the WFC3 and ACS fields (Section 4.1
and Figure~\ref{RGBmodel}) suggests an upper limit for any metallicity
gradient of $\sim -0.02$ dex/kpc, which would still leave the M101 halo
more metal-poor than most spiral galaxy halos studied to date. However,
this low metallicity is consistent with expectations from the trend
between halo metallicity and halo mass, given the estimates of a very
low halo mass for M101 (van Dokkum \etal 2014; Merritt \etal 2016).

Low mass stellar halos are also thought to have weaker metallicity
gradients. Expressed in terms of a \vmi\ color gradient in the RGB
populations, galaxies in the Harmsen sample had color gradients of
$\lesssim -0.004$ mag/kpc, with low mass halos showing little or no color
gradient at all. Our rough estimate of the maximum allowable RGB color
gradient between the ACS and WFC3 fields suggests a gradient of $-0.007$
mag/kpc, larger than any of the gradients seen in the Harmsen \etal
sample. Such a large gradient would seem indicative of a more massive
stellar halo (such as that in M31), at odds with the low halo mass
estimate for M101. However, a few considerations temper this result.
First and most importantly, the gradient we infer is an upper limit; the
true gradient may be significantly smaller. Second, Elias \etal (2018)
have used the Illustris simulations to argue that the non-spherical
shapes of galaxy halos can effectively reduce the observed halo surface
brightness profiles by as much as 2 \magsec\ in face-on spirals; if so,
the low M101 halo mass (inferred from surface photometry) may need to be
revised upwards and bring the inferred gradients in somewhat better
agreement with the relationship between halo mass and color gradient 
noted by Harmsen \etal (2017).

In summary, the low metallicity inferred for the M101 halo from the RGB
stars present in our WFC3 field is similar to that of the Milky Way's
outer halo, and broadly consistent with expectations for galaxies with
low halo mass and a quiet accretion history. For example, Elias \etal
(2018) used the Illustris simulations to investigate the behavior of
spirals with low stellar mass halos. In these simulations, the weak
stellar halos of galaxies like the Milky Way and M101 form early, and
are dominated by the accretion of low-mass satellites --- consistent with
the lower metallicity we see. Interestingly, these simulations also
predict that the M101 halo will have had a stronger {\sl in situ} halo
component inside of $r\sim 45$ kpc, which would have a higher
metallicity and lead to a stronger metallicity gradient between the
inner and outer halo. If so, this behavior may also explain the
relatively large gradient we infer for the halo metallicities between
the ACS and WFC3 fields.

\subsection{Comparing constraints from resolved populations and integrated light}

Our imaging data also give us the opportunity to do a direct comparison
of our results with those obtained from the broadband surface photometry
of M101's outskirts described in M+13. Integrated light surface
photometry has very different systematic uncertainties compared to
resolved stellar population studies --- surface photometry contends with
issues of sky subtraction, scattered light, and large scale
flat-fielding, while resolved population studies deal with uncertainties
due to background contamination, field crowding, and often limited field
of view. Thus, comparing results from studies conducted in the same
fields provide an important cross-check of systematic uncertainties
between the two techniques.

The ACS field was chosen specifically to lie on the extended ``Northeast
Plume" identified in broadband imaging by M+13. That study showed the
plume to be blue, with a total integrated $B-V$ color of $0.21 \pm
0.05$; coupled with the plume's faint luminosity and extremely low rate
of star formation (assessed by the far UV flux), M+13 inferred the
stellar populations in the plume formed during a weak burst of star
formation that peaked $\approx$ 250--350 Myr ago and has since largely
died out. The picture is well-supported by the Hubble data presented
here: the presence of the co-evol ``lump'' population seen in the ACS
CMD also argues for a fading burst population of roughly similar age.
However, the Hubble CMD sharpens the view significantly, revealing the
old RGB population as well as tracing the main sequence stars and helium
burning populations. On balance, the CMD modeling favors a slightly
older age ($\sim$ 300--400 Myr) for the starburst along with a more
slowly decaying star formation rate than the Gaussian model adopted in
M+13, but the CMD model that best matches the population mix also
matches the broadband color and surface brightness of the plume measured
over the ACS field of view (Table~\ref{IACmods}). The Hubble data also
breaks the age-metallicity degeneracy of broadband colors, revealing the
metal-poor nature of both the young and old stellar populations in the
field.

In contrast, the WFC3 pointing targeted a region where no broadband
light was detected in the M+13 imaging, putting an upper limit on the
surface brightness of $\mu_V > 29.0$ \magsec. In this region, the Hubble
CMD reveals an old stellar population, but with very low surface
density. Adopting a single burst stellar population model to match the
star counts measured along the RGB, we infer (after photometric
transformations, see Section 4.2) an integrated surface brightness of
$\mu_V = 30.86$ \magsec, nearly 2 \magsec\ fainter than the upper limit measured from
the broadband imaging of M+13. Thus the sparse halo population we detect
in the WFC3 field is consistent with the non-detection of broadband
light in the field by M+13.

Similarly deep imaging of M101 was used by van Dokkum \etal (2014) and
Merritt \etal (2016) to construct an {\sl azimuthally-averaged} surface
brightness profiles of M101 extending beyond the disk and into the
stellar halo. At the 23.3\arcmin\ distance of our WFC3 field, those
profiles show a surface brightness of $\mu_g \approx$ 32--33 \magsec.
Converting our inferred \mi\ surface brightness to $g$ yields
$\mu_g$=31.1 \magsec, or $\sim$ 1--2 \magsec\ {\sl brighter} than the
azimuthally-average profile. The source of this discrepancy remains
unclear. One possibility is substructure within M101's halo. Deep
imaging of many galaxies has revealed a wealth of accretion shells and
plumes in their stellar halos (\eg Martinez-Delgado \etal 2008, 2010;
Janowiecki \etal 2010; Ibata \etal 2014); if our WFC3 field has
serendipitously fallen on such a feature, we would infer a higher
surface brightness than the azimuthally-averaged value. At such depths,
where galaxy halos are thought to dominated by accretion, such
substructure could be significant (Bullock \& Johnston 2005, G\'omez
\etal 2013, Cooper \etal 2013), making azimuthal averaging highly
uncertain (see also the discussion in Watkins \etal 2016). Conversely,
the disagreement may simply be a sign of the systematic uncertainties
inherent to surface photometry at such faint light levels, where slight
differences in sky estimation and subtraction can lead to significant
changes in the extracted surface brightness profiles. With only one
pointing in M101's halo, it remains difficult to discriminate between
these possibilities.

\subsection{What happened to M101?}

Taken as a whole, the stellar populations in M101's outer disk provide
important constraints on M101's dynamical history. The galaxy's lopsided
disk and distorted outer isophotes have long suggested a recent
interaction with one of its neighboring companions. The high velocity HI
gas observed in both the NE Plume and the E Spur (van der Hulst \&
Sancisi 1988; Walter \etal 2008; Mihos \etal 2012) add to a picture in
which an encounter has driven a strong response in the galaxy's
morphology, kinematics, and star forming history. However, the identity
of the offending companion remains uncertain. One possibility is the
dwarf irregular galaxy NGC 5477, projected only 22\arcmin\ (44 kpc) away
from the center of M101. While its proximity is favorable for driving
a {\sl local} response in the galaxy's outer disk, it is difficult to
see how such a low luminosity dwarf could drive the strong {\sl global}
response of M101's asymmetric disk. A second candidate is the larger and
more luminous companion NGC 5474, located 44\arcmin\ (88 kpc) to the
south of M101. NGC 5474's lopsided morphology suggests it too has
suffered a recent interaction, and extremely diffuse HI is seen between
it and the larger M101 (Huchtmeier \& Witzel 1979, Mihos \etal 2012).
Nonetheless its distance from M101 and the lack of obvious tidal
features even at faint surface brightness (M+13) make the direct
connection to M101's dynamical history still somewhat circumstantial.

Now, however, the coeval burst population seen in the CMD of the outer
disk ACS field adds a new piece of evidence to the puzzle: the timescale
since interaction. If the ``lump'' stars formed as a response to a tidal
encounter, their age places the time since collision at $\Delta t
\approx$ 300--400 Myr. In loose groups, the typical encounter speed
involving galaxies is approximately that of a marginally bound orbit,
\ie $v_e \approx \sqrt{2} \times v_c$, where $v_c$ is the circular
velocity of the dominant galaxy, $\approx 200$ \kms\ for M101 (Bosma
\etal 1988). Therefore, the interacting companion would have traveled a
distance $v_e \times \Delta t \approx$ 100 kpc away from M101 in the
time since closest approach. In projection, the separation could be
significantly smaller, of course, but the rough agreement between the
expected separation and NGC 5474's projected distance from M101 (88 kpc)
lends support to a interaction scenario involving these two galaxies.
The timescale argument also works against scenarios involving the
smaller NGC 5477 --- while the dwarf is projected close to the NE Plume
{\sl today}, if the encounter happened 300 Myr ago as traced by the
burst population, then given M101's rotation speed, the outer disk has
gone through one-third of a rotation since the burst was triggered. The
observed proximity of NGC 5477 to the NE Plume may thus merely be
happenstance, with the dwarf promoting only a weak perturbation at best
on M101's already damaged outer disk. To test this scenario and fully
describe the dynamical evolution of the interactions between M101 and
its companion galaxies, detailed simulations are clearly warranted.

\section{Summary}

Using deep imaging from the ACS and WFC3 cameras on the {\sl Hubble
Space Telescope}, we have studied stellar populations in the outskirts
of the nearby spiral M101. We observed two fields: the ACS field
targeted M101's outer disk (cospatial with the Northeast Plume
discovered by Mihos \etal 2013) at a distance of 36 kpc from the
galaxy's nucleus, while the parallel WFC3 field was in an offset field
sampling the galaxy's outer stellar halo at a distance of 47 kpc. Our
photometry extends down to a limiting magnitude of \mi~=~28.6 (50\%
completeness limit), tracing RGB stars 3.5 magnitudes below the RGB tip,
and probing deep enough to detect upper main sequence stars with ages as
old as $\sim$ 400 Myr.

In the outer disk field probed by the ACS imaging, we find evidence of
multiple stellar populations indicative of a complex and extended star
formation history. In particular, we trace a population of evolved red
giant branch stars as well as young main sequence stars and blue and red
helium burning stars. The BHeB sequence shows a concentration of stars
at \Mi~=~$-2.4$ and \vmi~=~0.22, indicative of a co-evol stellar
population likely made in a recent weak burst of star formation in
M101's outer disk. Isochrone matching indicates an age of $\approx$
300--400 Myr and metallicity of [M/H]~$=-1.15 \pm 0.2$ for this
post-starburst population, while the older RGB stars appear to have a
slightly lower metallicity of [M/H]~$=-1.3 \pm 0.2$. These inferences
are supported by CMD modeling using the IACstar package, which also
shows that the burst population accounts for $\approx$ 10\% of the
stellar mass in the outer disk, and that the star formation rates have
declined by roughly half since the onset of the burst.

In the offset WFC3 halo field, the CMD shows an old RGB sequence with no
evidence for young stellar populations. The RGB in this field is $\sim$
0.1 mag bluer in \vmi\ than that observed in the ACS disk field,
implying a lower metallicity of [M/H]~$= -1.7 \pm 0.2$. We also see a
scattering of luminous stars on the AGB, suggesting the presence of an
intermediate age population. Comparisons with AGB populations found in
old, metal-poor stellar halos (Girardi \etal 2010) suggest this AGB
population is at least several Gyr old. The low metallicity of the M101
halo stars is similar to that of the Milky Way's outer halo, and
consistent with models where low mass stellar halos like M101's are
formed by early accretion events involving low mass satellites, with
very little accretion of massive objects at late times.

A comparison of these results from resolved stellar population work with
those derived from deep surface photometry provides an important test of
systematics between the two techniques. The stellar populations we
observed in the ACS field confirm the inferences of a recent burst of
star formation in M101's outer disk from our previous surface photometry
(M+13), but sharpen the view by revealing the presence of old RGB and
AGB stars, constraining the metallicities of both the young and old
stellar populations, and refining the recent evolution of the star
formation rate in the outer disk. The very low spatial density of halo
stars detected in the WFC3 field is consistent with the upper limit on
the broad band halo light in the field by M+13, but 1--2 \magsec\
brighter than {\sl azimuthally-averaged} profiles published by van
Dokkum \etal (2014) and Merritt \etal (2016). This discrepancy could be
a signature of substructure within M101's halo, or due to unresolved
systematics in the deep surface photometry.

In summary, we are left with a picture of M101 where a recent
interaction, likely with the companion galaxy NGC 5474, has stirred up
gas in the outer disk and triggered a weak burst of star formation
$\sim$ 300-400 Myr ago, which is now fading out. These types of
encounters are common in group environments (Sinha \& Holley-Bockelman
2012) and, by triggering these kinds of weak outer disk starburst
events, may lead to the slow growth of extended disks over time. At the
same time, though, the low metallicity and low mass of M101's stellar
halo argues that the galaxy has not experienced a significant number of
massive accretion events in its evolutionary history. Thus, in the loose
group environment where the dynamical timescales are long and mergers
less common, late-type spirals like M101 (and perhaps the Milky Way as
well) may walk an evolutionary path where their properties are shaped
more by weak encounters rather than the massive accretion events that
drive evolution in denser groups and clusters.

\acknowledgments

The authors would like to thank Andrew Dolphin for advice for the
DOLPHOT reductions, as well as Tom Brown and Miranda Link for their
assistance with the planning of the observations for this project. This
research made use of Astropy, a community-developed core Python package
for astronomy (Astropy Collaboration, 2013). JCM, PRD, and JJF were
supported through funds provided by NASA through grant GO-13701 from the
Space Telescope Science Institute, which is operated by the Association
of Universities for Research in Astronomy, Incorporated, under NASA
contract NAS5-26555.

Workload for this project: JCM and PRD led the observational design, PRD
led the data reduction and the analysis of the WFC3 halo field, JCM led
the analysis of the ACS disk field and the CMD modeling, and JJF led the
reduction and analysis of the Abell 2744 flanking field for background
estimation. PH and AEW contributed to the analysis and interpretation of
the data. JCM led the writing of this manuscript, with contributions
from all authors.

\facilities{HST, CWRU:Schmidt}

\software{DOLPHOT, Astropy}

\end{document}